\documentclass[pre,twocolumn]{revtex4-1}

\usepackage{graphicx}
\usepackage{amsmath}
\usepackage{mathtools}
\usepackage{amssymb}
\usepackage{pst-node}

\usepackage{verbatim}   % useful for program listings
\usepackage{color}      % use if color is used in text
\usepackage{subfigure}  % use for side-by-side figures
\usepackage{hyperref}   % use for hypertext links, including those to external documents and URLs

\usepackage{braket}

\newcommand{\ba}{\begin{eqnarray}}
\newcommand{\ea}{\end{eqnarray}}
\newcommand{\be}{\begin{equation}}
\newcommand{\ee}{\end{equation}}

\begin{document}

%==============================================================
\title{Thermodynamic collapse in a lattice-gas model for a two-component system of penetrable 
particles}
%for a two-component mixture} 
%==============================================================

%\author{Daniel Alejandro G\'omez Bustos}
%\affiliation{Department of Chemistry, Federico Santa Maria Technical University, Campus San Joaquin, 
%Santiago, Chile}
\author{Derek Frydel}
\affiliation{Department of Chemistry, Federico Santa Maria Technical University, Campus San Joaquin, Santiago, Chile}
\author{Yan Levin}
\affiliation{Institute of Physics, The Federal University of Rio Grande do Sul, Porto Alegre 91501-970, Brazil}

\date{\today}

\begin{abstract}
We study a lattice-gas model of penetrable particles on a square-lattice substrate with same-site and 
nearest-neighbor interactions.  Penetrability implies that the number of particles occupying a single 
lattice site is unlimited and the model itself is intended as a simple representation of penetrable particles 
encountered in realistic soft-matter systems.  Our specific focus is on a binary mixture, where particles of 
the same species repel and those of the opposite species attract each other.  As a consequence of 
penetrability and the unlimited occupation of each site, 
the system exhibits thermodynamic collapse, which in simulations is manifested by an 
emergence of extremely dense clusters scattered throughout the system with energy 
of a cluster $E\propto -n^2$ where $n$ is the number of particles in a cluster.  
After transforming a particle system into a spin system, in the large density 
limit the Hamiltonian recovers a simple harmonic form, resulting   
in the discrete Gaussian model used in the past to model the roughening transition of 
interfaces.  For finite densities, due to the presence of a non-harmonic term, the system is 
approximated using a variational Gaussian model.  
\end{abstract}

\pacs{
}

\maketitle
%------------------------------------------------

\section{Introduction}

In a recent article \cite{Frydel18b} we studied a one-dimensional lattice gas model of penetrable 
particles and demonstrated that a two-component system (where particles of the same species repel 
and those of opposite species attract each other) becomes thermodynamically unstable, where the 
collapsed state is manifested by the presence of scattered and extremely dense clusters, in which 
the occupation number of a site that is part of the cluster is $n\gg 1$.  
This behavior is not unique to lattice models and has been previously observed in more realistic systems 
of penetrable particles such as a penetrable sphere model \cite{Frydel16,Frydel17,Frydel18a}.  
Prior to these examples, the possibility of thermodynamic collapse in a multicomponent 
system of soft particles 
has been considered as early as 1966  by Ruelle and Fisher \cite{Ruelle66a,Ruelle66b,Heyes07}, 
who also explored mathematical 
criteria for the conditions in which such a collapse becomes plausible.  

The renewed interest in penetrable particles has been triggered by a growing number of synthesized 
and naturally occurring nanoparticle whose pair interactions lack the usual hard-core repulsion, resulting 
in ultrasoft particles that interpenetrate and, in principle, can occupy the same space \cite{Likos01a}.  
Penetrability gives rise to different behaviors than those encountered in systems with hard-core repulsion.  
The type of soft interactions, furthermore, plays a decisive role in determining a particular 
behavior of the system \cite{Likos01}.

In a one-component system, thermodynamic collapse becomes possible for systems with 
pair interactions comprised of a short-range attractive tail and a repulsive soft-core.  
More recent examples where such systems are studies in connection to thermodynamic
collapse include Ref. \cite{Malescio15,Malescio16,Malescio18}, among others.  
The most famous example of thermodynamic collapse, however, is that in gravitational system 
\cite{Yan14}, whose pair interaction consists of only attractive long-range part.  
When it comes to two-component systems, a considerably less work has been done 
to understand the mechanism of thermodynamic collapse.

Thermodynamic collapse in a two-component system is not self-evident, since attractive interactions 
occur between particles of opposite species, and this implies that a collapsed configuration, or a 
group of configurations, involves a very specific arrangement of particles whose specific structures 
has been investigated in Ref. \cite{Frydel18b} for a one-dimensional lattice-gas model.

Because one-dimensional models, as a general rule, preclude the possibility of a phase transition 
\cite{Cuesta04} (interestingly enough, this rule does not apply to thermodynamic collapse), the 
investigation in the Ref. \cite{Frydel18b} is not entirely satisfactory.  In the present article we consider 
a binary system on a lattice-square substrate with nearest neighbor interactions, as it is the most 
standard model in two-dimensions.   Because the occupation number is unlimited, the system 
is closely related to the discrete Gaussian model originally designed to capture the structure and 
behavior of interfaces and the roughening transition \cite{Chui76,Weeks80,Binder95},

Our results are organized as follows.  In Sec. \ref{sec:model} we introduce 
the model and write down the corresponding grand partition function.  In this section 
we introduce two distinct ways of counting particles, depending on whether particles are 
considered as distinguishable or indistinguishable.  Different ways of counting particles 
does not arise for a single occupation lattice-gas models and is a consequence of 
multiple occupation.  
In Sec. \ref{sec:trans} we transform the original particle system into spin ensemble.  
In the transformed ensemble 
spins can take on any integer value as a consequence of particle 
penetrability.   
In Sec. \ref{sec:infty} we analyze thermodynamic collapse 
in the infinite density limit.  
%The limit is the 
This limit is the 
consequence of penetrability 
and implies that the average occupation of a site is $n\to\infty$.  
%where $n$ is the occupation number of a site.  
In this limit the Hamiltonian reduces to 
a harmonic function and the corresponding partition function transforms into a discrete 
Gaussian model (DG).  
A similar model was used to study roughening transition of interfaces. 
%which allows us to draw connection between
% the thermodynamic collapse in our model and the roughening transition of the
% interface model.  
In Sec. \ref{sec:rho} we analyze the system at finite density.  Due to a non-harmonic 
term, the resulting partition function is no longer Gaussian and we analyze the system 
using a Gaussian variational method.  Both approximate and exact models indicate 
the presence of a metastable region, so that even though the global minimum 
corresponds to a collapsed state, the system remains in 
metastable equilibrium.

\section{The model}
\label{sec:model}

The model consists of two types of particles on a two-dimensional square-lattice substrate.  
As hard-core interactions are not included, there is no restriction on the number of particles 
that can occupy a single site.  If the occupation numbers for a given site are $n_{i}^+$ and 
$n_{i}^-$, where the superscripts ``+'' and ``-'' designates different species, then the 
Hamiltonian of the system is
\ba
H&=& K \sum_{} \bigg[\frac{1}{2} n_{i}^+(n_{i}^+ - 1) + \frac{1}{2} n_{i}^-(n_{i}^- - 1) - n_{i}^+n_{i}^-\bigg] 
\nonumber\\ 
&+& \alpha K \sum_{nn} \bigg[n_{i}^{+}n_{j}^{+} +  n_{i}^{-}n_{j}^{-}  - n_{i}^+n_{j}^-  - n_{i}^-n_{j}^+\bigg],  
\nonumber\\ 
\label{eq:H1}
\ea
where the first line is for the interaction between particles on the same site, and the second 
line is for the interaction between particles on neighboring sites (the subscript $nn$ indicates 
the nearest-neighbor interaction).  The dimensionless coupling parameter $\alpha$ for interactions 
between neighbors is positive in our model.  This implies that particles of opposite species 
attract and those of the same species repel each other.

The fact that each lattice site can be occupied by multiple particles at one time results in two 
types of statistics.  If particles are distinguishable as in classical fluids, then the grand 
canonical partition function is 
\be
\Xi_{a} = 
\sum_{n_1^+=0}^{\infty} \sum_{n_1^-=0}^{\infty}\dots \sum_{n_N^+=0}^{\infty} \sum_{n_N^-=0}^{\infty}
e^{-\beta H_{\rm int}} \prod_{i=1}^N \frac{e^{\beta \mu' (n_i^++n_i^-)}}{n^+_i!n^-_i!},
\label{eq:X_dist}
\ee
where   
\be
H_{\rm int} = \frac{K}{2} \sum_{} \big(n_{i}^{+}-n_{i}^-\big)^2 
+ \alpha K \sum_{nn} (n_{i}^+-n_{i}^-) (n_{j}^+ - n_{j}^-) 
\ee
is the interaction Hamiltonian, 
\be
\mu' = \mu + \frac{K}{2}
\ee
is the effective chemical potential, and $N=L^2$ is the number of lattice sites, where $L$ is the 
size of the system.  The factor $1/n_i!$, also referred to as the Gibbs correction, is a feature
of distinguishable particles, and indicates that statistics at a single site follows a poisson 
rather than an exponential distribution.  A more detailed analysis of distinguishability versus
indistinguishability is provided in Ref. \cite{Frydel18b}.

On the other hand, if particles are regarded as indistinguishable, a situation which in classical 
systems arises for example in growth models, where particles do not change their location on 
the lattice substrate but rather are added or removed from it at each Monte Carlo step, in which 
case the particles of a given site have no labels, then the grand partition function is  
\be
\Xi_{b} = 
\sum_{n_1^+=0}^{\infty} \sum_{n_1^-=0}^{\infty}\dots \sum_{n_N^+=0}^{\infty} \sum_{n_N^-=0}^{\infty}
e^{-\beta H_{\rm int}} \prod_{i=1}^N e^{\beta \mu' (n_i^++n_i^-)}.
\label{eq:X_indist}
\ee

Based on the above discussion, even if the systems obey the same Hamiltonian, they can be subject to 
different rules of statistical mechanics which, in turn, can lead to different behaviors.  
This difference can be particularly relevant in characterizing thermodynamic collapse.  As this issue
does not arise in a standard lattice-gas model with occupations limited to one, it is important to 
emphasize it as well as consider it in overall analysis.

\section{transformation into a spin-ensemble}
\label{sec:trans}

The system described above can be simplified by transforming it into a spin ensemble 
with spins corresponding to $s_i=n_i^+-n_i^-$.  
Because a single configuration in the spin-ensemble corresponds to infinitely many 
configurations in 
the particle-ensemble, these degeneracies need to be correctly accounted for.  The resulting 
transformed partition functions are \cite{Frydel18b}
\be
\Xi_{a} = 
\!\!\!\! \sum_{s_1=-\infty}^{\infty} \!\!\! \dots  \!\!\! \sum_{s_N=-\infty}^{\infty}   
\!\!\! e^{- \beta K \alpha \sum_{nn}s_is_{j}}  \prod_{i=1}^N 
\bigg[ e^{-\frac{\beta K}{2} s_i^2}  {\rm I}_{s_i} \big(2e^{\beta \mu'}\big)\bigg], 
\label{eq:Xs_dist}
\ee
and 
\be
\Xi_{b} = 
\!\!\!\! \sum_{s_1=-\infty}^{\infty} \!\!\! \dots  \!\!\! \sum_{s_N=-\infty}^{\infty}   
\!\!\! e^{- \beta K \alpha \sum_{nn}s_is_{j}} \prod_{i=1}^N
\bigg[ e^{-\frac{\beta K}{2} s_i^2}  \frac{e^{\beta \mu'|s_i|}}{1-e^{2\beta\mu'}}\bigg], 
\label{eq:Xs_indist}
\ee
for distinguishable and indistinguishable particles, respectively.  The terms inside square
brackets can be regarded as effective external field.  Furthermore, as these terms are even 
function in $s_i$, the spin symmetry is never broken so that $\langle s_i\rangle=0$
under all conditions.  The 
function ${\rm I}_s(x)$ in Eq. (\ref{eq:Xs_dist}) is the modified Bessel function of the first 
kind. 

Any quantity defined in the original ensemble can be calculated as another quantity 
in the spin-ensemble.  For example, the average number of particles at a single site $i$, 
in the original ensemble defined as 
\be
\rho_i = \langle n_i^+\rangle +\langle n_i^-\rangle = \frac{1}{N}\frac{\partial\ln \Xi}{\partial\beta\mu}, 
\label{eq:rho}
\ee
in the spin-ensemble becomes 
\be
\rho_i = e^{\beta \mu'}
\Bigg\langle \frac{{\rm I}_{s_i+1} \big(2e^{\beta\mu'}\big)+{\rm I}_{s_i-1} 
\big(2e^{\beta \mu'}\big)}{{\rm I}_{s_i} \big(2e^{\beta \mu'}\big)} \Bigg\rangle_s, 
\label{eq:rho2a}
\ee
for distinguishable particles, where the subscript $s$ indicates the average calculated 
in the spin ensemble, and 
\be
\rho_i = \langle |s_i| \rangle_s + \frac{2e^{2\beta \mu'}}{1-e^{2\beta \mu'}}, 
\label{eq:rho2b}
\ee
for indistinguishable particles.  For distinguishable particles, 
the limit $\rho_i\to\infty$ is attained if $\mu'\to\infty$, and for indistinguishable particles  if $\mu'\to 0^-$.   
In the rest of the paper, we use $\rho\equiv \rho_i$, to indicate the average number of particles on any 
lattice site and refer to $\rho$ as density.  
The limit $\rho\to\infty$ is a consequence of the fact that no limit is placed on the occupation 
number.  This is quite different from the standard lattice-gas model where the maximum 
density is $\rho=1$.

%\subsection{analysis}
The spin-ensembles in Eq. (\ref{eq:Xs_dist}) and Eq. (\ref{eq:Xs_indist}) more generally can be 
written as 
\be
\Xi = B^N\sum_{s_1=-\infty}^{\infty} \!\! \dots  \!\! \sum_{s_N=-\infty}^{\infty}e^{-\beta H}, 
\ee
with the pre-factors 
\be
  B(\mu') =
  \begin{cases}
    {\rm I}_{0} (2e^{\beta\mu'}),~~\text{distinguishable} \\
    \frac{1}{1-e^{2\beta\mu'}},~~~~\text{indistinguishable}, 
  \end{cases}
  \label{eq:B}
\ee
and the Hamiltonian is given by 
\be
H = \alpha K \sum_{nn} s_i s_j + \frac{K}{2}\sum s_i^2 + \sum h(s_i).  
\label{eq:H_s}
\ee
where the one-body potentials $h(s_i)$ are
\be
  \beta h(s_i) =
  \begin{cases}
     -\ln\Big[\frac{{\rm I}_{s_i} (2e^{\beta\mu'})}{{\rm I}_{0} (2e^{\beta\mu'})}\Big],~~\text{distinguishable}  \\
      -\beta \mu' |s_i|,~~~~~~~~~\text{indistinguishable}.  
  \end{cases}
  \label{eq:hs}
\ee

\begin{figure}[h] 
 \begin{center}
 \begin{tabular}{rrrr}
  \includegraphics[height=0.18\textwidth,width=0.22\textwidth]{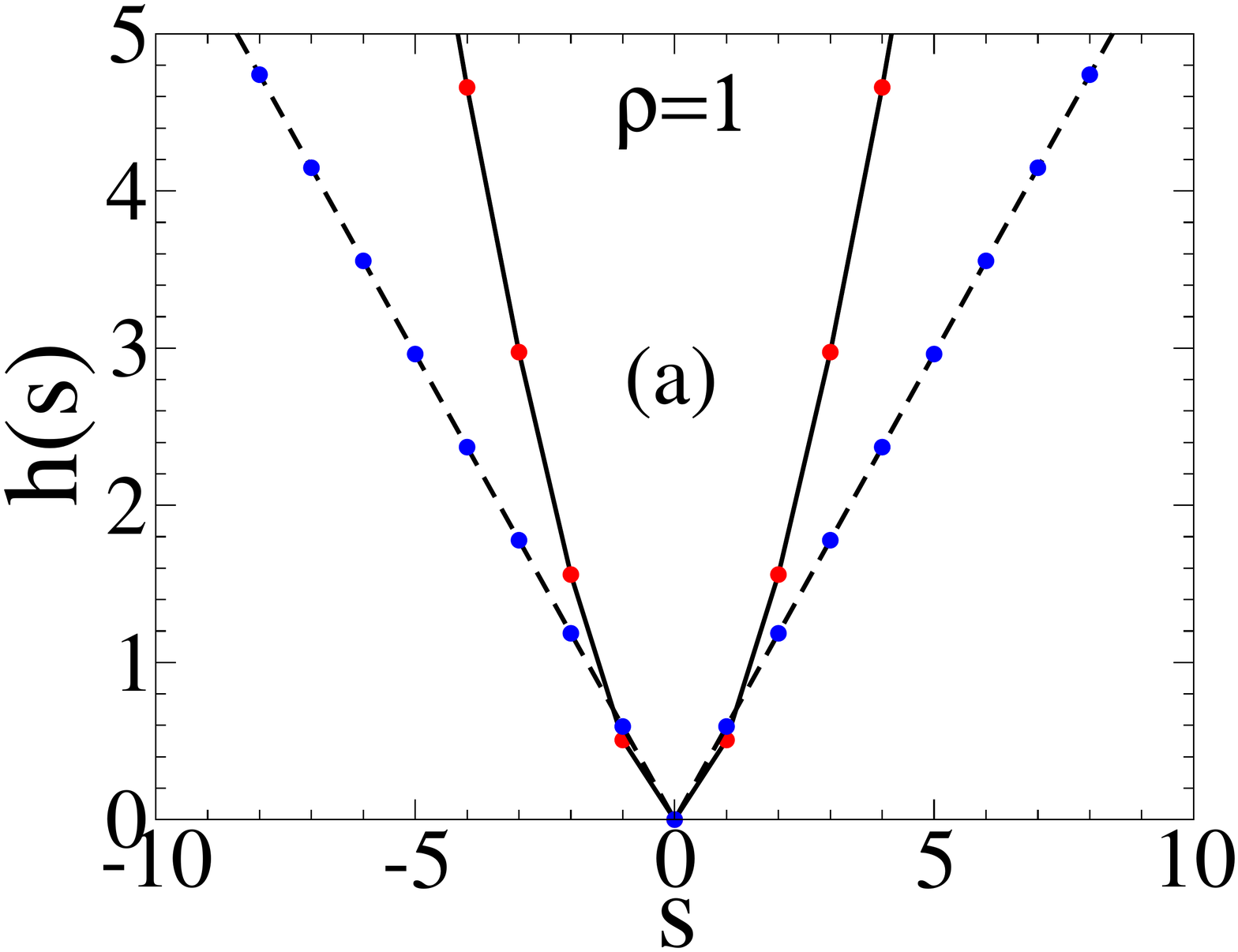}&
  \includegraphics[height=0.18\textwidth,width=0.22\textwidth]{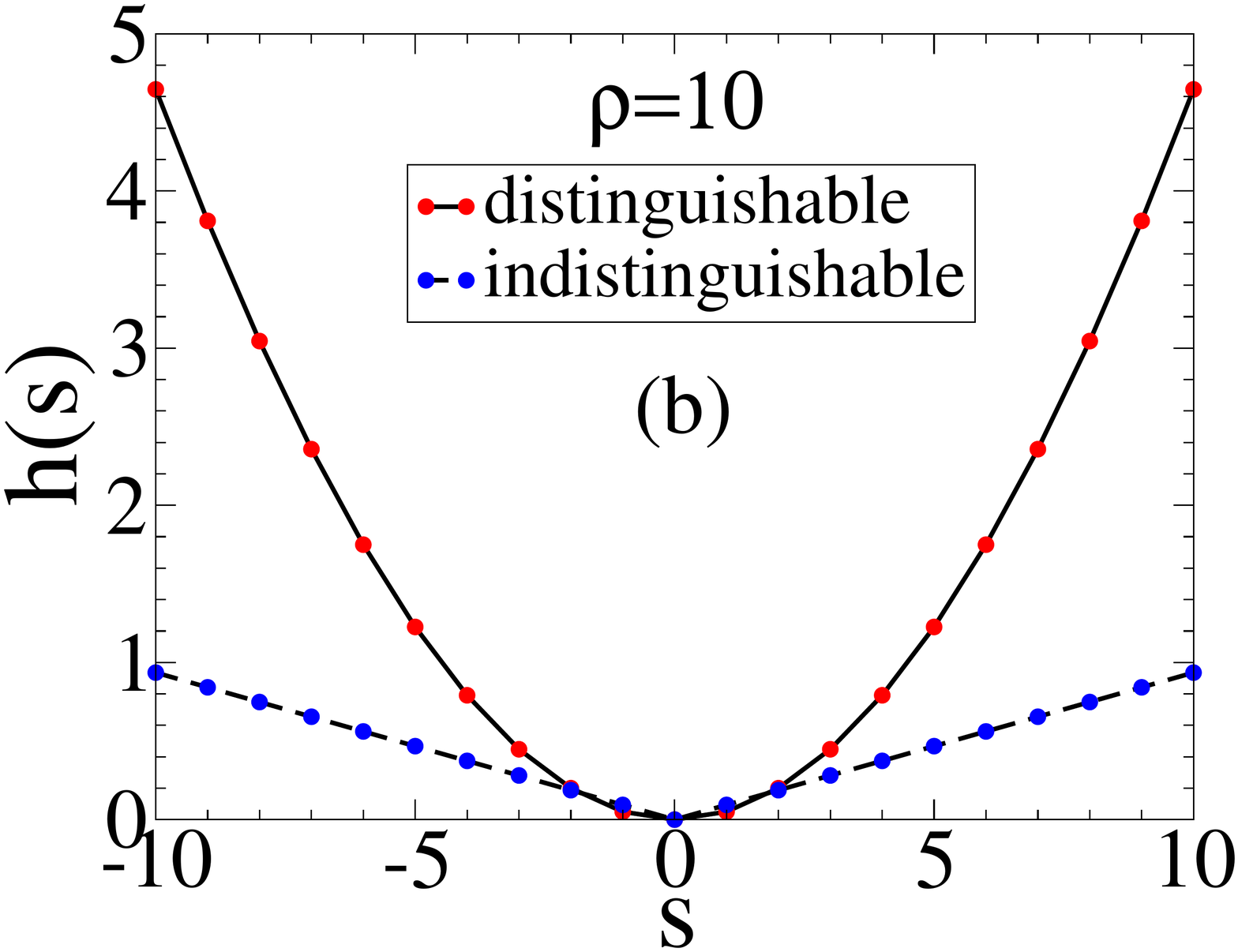}\\
 \end{tabular}
 \end{center}
\caption{The functions $h(s)$ for distinguishable and indistinguishable particles for 
$\beta K=5$ and $\alpha=1/4$ and for two different densities, see  
Eq. (\ref{eq:hs}).  For $\rho=1$ in (a), $\beta\mu'=-0.27$ and $\beta \mu'=-0.57$, and 
for $\rho=10$ in (b), $\beta \mu'=1.66$ and $\beta \mu'=-0.09$  for 
distinguishable and indistinguishable particles, respectively.  }
\label{fig:h} 
\end{figure}

Note that in the limit $\rho\to \infty$, $h(s_i)\to 0$ and both Hamiltonians 
become a simple harmonic function.  The difference between distinguishable and 
indistinguishable particles, therefore, becomes relevant at finite densities.   
For illustration and to see how these differences might be manifested, 
in Fig. (\ref{fig:h}) we plot $h(s)$ for distinguishable and indistinguishable particles for 
the parameters $\beta K=5$ and $\alpha=1/4$.  Based on the figure, 
one may expect larger fluctuations for indistinguishable particles due to the 
shape of the function $h(s_i)$.  
%For indistinguishable particles $h(s)$ is linear, and for distinguishable particles it is can approximately be regarded as harmonic.  
%%%%%%%%%%%%%%%%%%%%%%
\graphicspath{{figures/}}
%%%%%%%%%%%%%%%%%%%%%%%

\subsection{connection with other spin models}

It might be of interest to place our spin model in the context of other related models.  
The first difference to be noted is that 
unlike the standard Ising model, our model permits a spin $s_i=0$, which can be 
regarded as an empty site.  The class of Ising models that permit empty sites are referred to 
as site-diluted Ising models with the Hamiltonian 
$H_{} = -J \sum_{nn} p_ip_j s_i s_j$, where $s_i=\pm 1$ and $p_i=0,1$ are random (correlation free) 
occupation numbers such that $\langle p_i\rangle=\rho$ \cite{Parisi97,Rosinberg99}.  
These models assume the presence of defects in the lattice structure in a magnetic material 
and represent quenched dilution.  Models describing 
annealed dilution are possible and have been studied in the past \cite{Romano07}.  
%One notable feature of dilute Ising models is that they exhibit crossover to percolation behavior.  
Our model can be regarded as a version of a site-diluted (annealed) model, which would be interesting to 
study in its own right by limiting spins to $s_i=-1,0,1$, where the frequency of empty spins is 
determined by the function $h(s_i)$.  

Our model bears the closest analogy to the discrete Gaussian model (DG) \cite{Chui76,Sly16} 
dubbed so by Chui and Weeks in 1976.  The DG model belongs to a family of random surface 
models and whose Hamiltonian is given by $H_{} = \frac{1}{2}J \sum_{nn} (s_i-s_j)^2 + 4hJ \sum_{n} s_i^2$.  
In the limit $\rho\to\infty$, where $h(s)=0$, our model corresponds to the DG model.  For 
the parameter $h=0$, the DG model can be mapped onto a lattice Coulomb system, 
and like the lattice Coulomb model, it exhibits the Kosterlitz-Thouless transition.   This corresponds
to our parameter $\alpha=1/4$.  In appendix \ref{sec:A3}

\subsection{simulation details}
In addition to analytical results, we study the transformed spin ensemble using Monte Carlo
simulation.  
%The simulation details are as follows.  
%we next show simulation results, we briefly review details of the algorithm.  
The simulated system consists of spins on a square-lattice substrate.  
A simulation box itself is a square of size $L=128$ with periodic boundary conditions.  
A Monte Carlo move consists of a random selection 
of a lattice site followed by the trial change of the spin by either $1$ or $-1$ with 
equal probability.  The move is accepted if it lowers the energy, otherwise it is accepted 
with the probability $e^{-\beta (H_{new}-H_{old})}$.  Before calculating average quantities,
the system is equilibrated for half a million steps.  The average quantities are subsequently 
computed during another $2$ million steps.

\section{The limit $\rho\to \infty$}
\label{sec:infty} 

In the limit $\rho\to\infty$, $h(s_i)$ as defined in Eq. (\ref{eq:hs}) vanishes 
and the Hamiltonian in Eq. (\ref{eq:H_s}) for both distinguishable and indistinguishable
particles attains a simple quadratic form 
\be
H_{\infty} = \alpha K \sum_{nn} s_i s_j + \frac{K}{2}\sum s_i^2, 
\label{eq:H_infty}
\ee
whose Boltzmann factor is a Gaussian function and, as the spins are restricted to integers, the 
resulting system is a discrete Gaussian model (DG).  In the past, the DG model has been used
to model an interface
%where it captures roughening transition 
\cite{Chui76,Weeks80,Binder95}.  
Although the interpretation and the parametrization of that DG model for interfaces is different 
from ours (in the interface model spins represent height of an interface and, as the heights of 
neighboring spins tend to be the same, $\alpha<0$), the same general analysis applies to both.  
The analogy between the interface model and the present binary lattice-gas system of penetrable 
particles is also interesting.  
%The roughening transition in the former 
%is interpreted as the thermodynamic collapse in the latter.  

Even though the partition function of the DG model has a Gaussian form, it cannot be solved exactly. 
%as in the case of a continuous Gaussian model (CG) where spins are continuous \cite{Moshe14,Mattis06}.  
However, if we neglect spin discreteness, it may be possible to approximate the DG model 
with the continuous Gaussian model (CG) which can be solved 
exactly \cite{Moshe14,Mattis06}.    

%But by ignoring the discreteness of the spins, it is not clear what information will actually be 
%lost.  To be able to see , we follow the 

%In 1976 Chui and Weeks considered the DG model as a simple representation of an interface 
%\cite{Chui76,Weeks80,Binder95}.  In the interface interpretation of the DG model spins represent 
%an interface height at a lattice site, and to capture the tendency of neighboring sites to have similar 
%height, the coupling between the sites is negative, $\alpha<0$.  

A systematic way to carry this out is to write the partition function for the DG model 
where the partition function of the CG model is a contributing term.  Any additional term
would then represent contributions due to spin discreteness.  
%We proceed in a way that will allow us to separate the partition function for the DG model 
%into the CG counterpart plus the remaining contributions.  
%The way to proceed
%by separating the partition function of the DG model into the CG model plist contributions.  
To see if this can be done, we first reformulate the Hamiltonian in Eq. (\ref{eq:H_infty}) using matrix 
notation, 
\be
H_{\infty} = \frac{K}{2}\sum_{i,j} A_{ij}s_is_j = \frac{K}{2} {s}^T{A}{s}, 
\label{eq:H_infty_B}
\ee
where ${s}=(s_1,\dots,s_N)$ is the $N$-dimensional vector, ${A}$ is a $N\times N$ matrix 
with elements 
\be
A_{ij} = \delta_{ij} + \alpha \epsilon_{ij}, 
\ee
where $\delta_{ij}$ is the Kronecker delta function, and $\epsilon_{ij}=1$ if the two spins 
are the nearest neighbors and zero otherwise.  ${A}$ for an arbitrary dimension $d$ is 
given in Appendix (\ref{sec:A1}).  The corresponding partition function is
\be
\Xi_{\infty} = \sum_{s_1=-\infty}^{\infty} \!\! \dots  \!\! \sum_{s_N=-\infty}^{\infty}
e^{-\frac{\beta K}{2} {s}^T{A}{s}}.  
\label{eq:Xi_infty}
\ee
Note that we ignore the pre-factor $B$ defined in Eq. (\ref{eq:B}) which in the limit 
$\rho\to\infty$ diverges, however, regardless of its value, it does not affect configurations.

If we rewrite the partition function in Eq. (\ref{eq:Xi_infty}) as 
\be
\Xi_{\infty} =  
\prod_{i=1}^N \int_{-\infty}^{\infty} ds_i \sum_{n_i=-\infty}^{\infty} \delta(s_i-n_i)\,\,
e^{-\frac{\beta K}{2} {s}^T{A}{s}}, 
\ee
and express the Dirac comb function as a Fourier series, 
\be
\sum_{n=-\infty}^{\infty} \delta(s-n) = \sum_{k=-\infty}^{\infty} e^{i2\pi k s},
\ee
we arrive at
\ba
\Xi_{\infty} &=& \sum_{k_1=-\infty}^{\infty}\dots\sum_{k_N=-\infty}^{\infty} \nonumber\\
&\times& \bigg[
\int_{-\infty}^{\infty} d {s_1}\dots\int_{-\infty}^{\infty} d {s_N}\,e^{i2\pi {\bf k}\cdot{\bf s}}
e^{-\frac{\beta K}{2} {s}^T{A}{s}}\bigg], 
\ea
where the integral term in square brackets is a Gaussian integral with a linear term that  
can be evaluated exactly using the identity
\be
\int d{\bf x}\, e^{i {\bf k}\cdot{\bf s}} 
e^{-\frac{1}{2} {s}^T{A}{s}} =  
e^{-\frac{1}{2} {k}^T {A}^{-1} { k}} \sqrt{\frac{(2\pi)^N}{\det { A}}}, 
\ee
where $A^{-1}$ is the inverse of the matrix $A$.  
The resulting partition function is comprised of two subsystems, 
\be
\Xi_{\infty} = \Xi_G\, \Xi_L, 
\label{eq:Xi_gen}
\ee
where $\Xi_G$ is the partition function of the CG model,  
\be
\Xi_G = \bigg(\frac{2\pi}{\beta K}\bigg)^{N/2}  \sqrt{\frac{1}{\det {A}}}, 
\label{eq:Xi_G}
\ee
and $\Xi_L$ represents all the contributions due to spin discreetness and is given by 
\be
\Xi_L = \sum_{s_1=-\infty}^{\infty}\dots\sum_{s_N=-\infty}^{\infty} 
e^{-\frac{1}{2} \frac{1}{\beta K/(4\pi^2)} ~ {s}^T {A}^{-1} {s}},
\label{eq:Xi_L}
\ee
%The isomorphism between $\Xi_L$ and $\Xi_{\infty}$   
%establishes duality relation between the two models.  It relates the free energy of one model 
%at low temperature to that of another model at high temperature.  
The dimensionless temperature of $\Xi_L$ is $k_BT' = \beta K/(4\pi^2)$.

\subsection{continuous Gaussian model}

From Eq. (\ref{eq:Xi_gen}) it is seen that by approximating the DG model as 
$$
\Xi_{\infty} \approx \Xi_G, 
$$
the missing contributions due to the spin discreteness are contained in the term $\Xi_L$.  
In this section we verify how accurate this approximation is.  To do this, we 
need to evaluate $\Xi_G$.

%To proceed with the solution to $\Xi_{\infty}$, we deal with the continuous Gaussian counterpart $\Xi_G$ first.  
The determinant in Eq. (\ref{eq:Xi_G}) is solved using the identity 
\be
\det{A} = \prod_{k=1}^N\lambda_k,
\ee 
where $\lambda_k$ are the eigenvalues of $A$.  $A$ is a circulant block matrix  with 
circulant blocks \cite{Davis79,Chen87,Kaveh11}.  The eigenvalues of a circulant matrix 
are Fourier modes.  For a matrix $A$ in $d=2$ the eigenvalues are 
\be
\lambda(q_1,q_2) = 1 + 2\alpha \cos q_1 + 2\alpha \cos q_2,
\label{eq:lambda}
\ee
where 
\be
q_i = \frac{2\pi n_i}{L}, ~~~ n_i=0,1,\dots,L-1
\ee
so that in total there are $N=L^2$ eigenvalues.  The determinant of $A$ now becomes 
\be
\det{A} =  e^{\sum_{n_1=0}^{L-1}\sum_{n_2=0}^{L-1}\ln\big[1 + 2\alpha \cos(\frac{2\pi}{N}n_1) + 2\alpha \cos(\frac{2\pi}{N}n_2)\big]}, 
\ee
which in the thermodynamic limit $L\to\infty$ becomes 
\be
\det{A} =  e^{(\frac{L}{2\pi})^2 \int_0^{2\pi} dq_1 \int_0^{2\pi} dq_2\, \ln[1 + 2\alpha \cos q_1 +  2\alpha \cos q_2]}.  
\ee
To complete the expression, it remains to evaluate the integral
\be
I = \bigg(\frac{1}{2\pi}\bigg)^2 \!\! \int_0^{2\pi} \!\! dq_1 \int_0^{2\pi} \!\! dq_2\, \ln\big[1 + 2\alpha \cos q_1 +  2\alpha \cos q_2\big].  
\label{eq:I}
\ee
When evaluated, it corresponds to a hypergeometric function which can also be expressed as 
a power series in $\alpha$, 
\be
I = -\sum_{k=1}^{\infty} \frac{\alpha^{2k}}{2k} \frac{(2k)!^2}{k!^4}.  
\ee
The interval of convergence of the above series is $|\alpha|\le 1/4$.  At $\alpha=1/4$, $I$ 
remains finite with a value $I\approx -0.220$.  For any value outside the radius of convergence, 
the series diverges, which in the present model implies thermodynamic instability.  We designate 
this value of $\alpha$ as $\alpha_c$.  

Given the above results, the partition function in Eq. (\ref{eq:Xi_G}) becomes 
\be
\Xi_G = \bigg(\frac{2\pi}{\beta K}\bigg)^{N/2} \exp\bigg[\frac{N}{4}\sum_{k=1}^{\infty} \frac{\alpha^{2k}}{k} \frac{(2k)!^2}{k!^4}\bigg].  
\label{eq:Xi_G2}
\ee

It is interesting to consider at this point the partition function of the Ising model 
that can be expressed as (see appendix \ref{sec:A2})
\be
\Xi_{IS} = [2\cosh(2\beta J)]^N \exp\bigg[-\frac{N}{4}\sum_{k=1}^{\infty} \frac{\alpha^{2k}}{k} \frac{(2k)!^2}{k!^4}\bigg]
\ee
where $\alpha$ is a function of $\beta J$ according to 
\be
\alpha = \frac{1}{2}\frac{\sinh(2\beta J)}{\cosh^2(2\beta J)},
\label{eq:alpha_is}
\ee
and $J$ is the interaction strength between nearest neighbor sites.  
In both the DG and the Ising model the value $\alpha=1/4$ has physical significance.  In the Ising 
model it indicates a critical point of a continuous phase transition and in the Gaussian model it is 
the last point before thermodynamic instability.  The Ising model, however, is prevented from
leaving the convergence region as a result of the parametrization in Eq. (\ref{eq:alpha_is}), 
and thermodynamic instability never precipitates.

Going back to the partition function $\Xi_G$, we point out that even if $\Xi_G$ is finite at $\alpha_c$
other quantities may diverge.  The internal energy defined as 
\be
\beta u = -\frac{\alpha}{N} \frac{\partial \log \Xi_G}{\partial \alpha} = 2\alpha K \langle s_is_j\rangle, 
\label{eq:u0}
\ee
where $\langle s_is_j\rangle$ are spin correlations between two nearest neighbors, 
can be calculated exactly using Eq. (\ref{eq:Xi_G2}), leading to 
\be
\beta u(\alpha) = \frac{1}{2} - \frac{1}{\pi} {\rm K}\big(16\alpha^2\big)
\label{eq:u}
\ee
where $ {\rm K}(x)$ is the complete elliptic integral of the first kind, which contains 
logarithmic singularity at $\alpha_c$, 
\be
\beta u(\alpha) \approx   \frac{1}{2} + \frac{1}{2\pi}\ln\bigg(\frac{1-4|\alpha|}{8}\bigg).  
\ee

\begin{figure}[h] 
 \begin{center}
 \begin{tabular}{rrrr}
  \includegraphics[height=0.18\textwidth,width=0.22\textwidth]{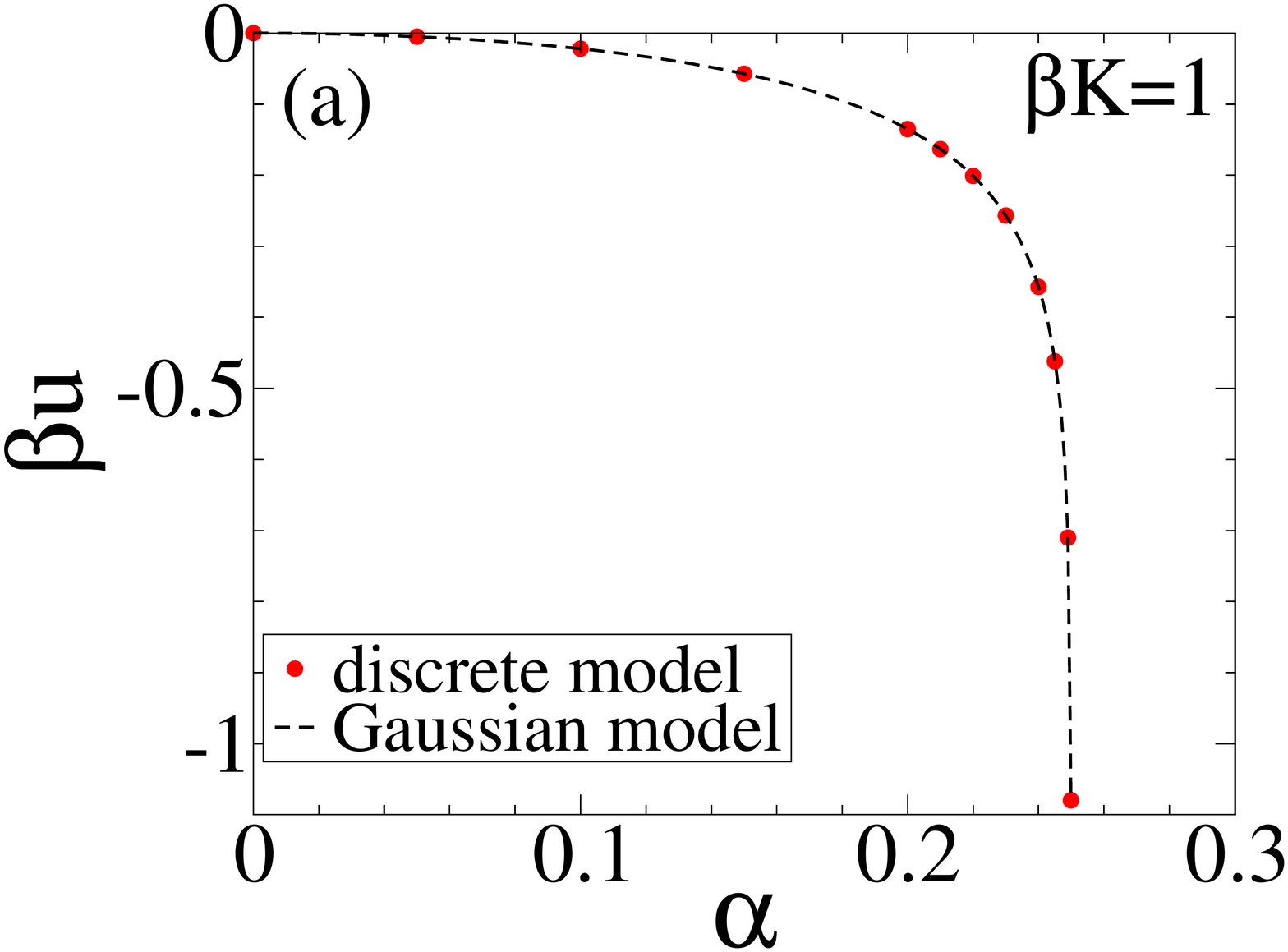}&
  \includegraphics[height=0.18\textwidth,width=0.22\textwidth]{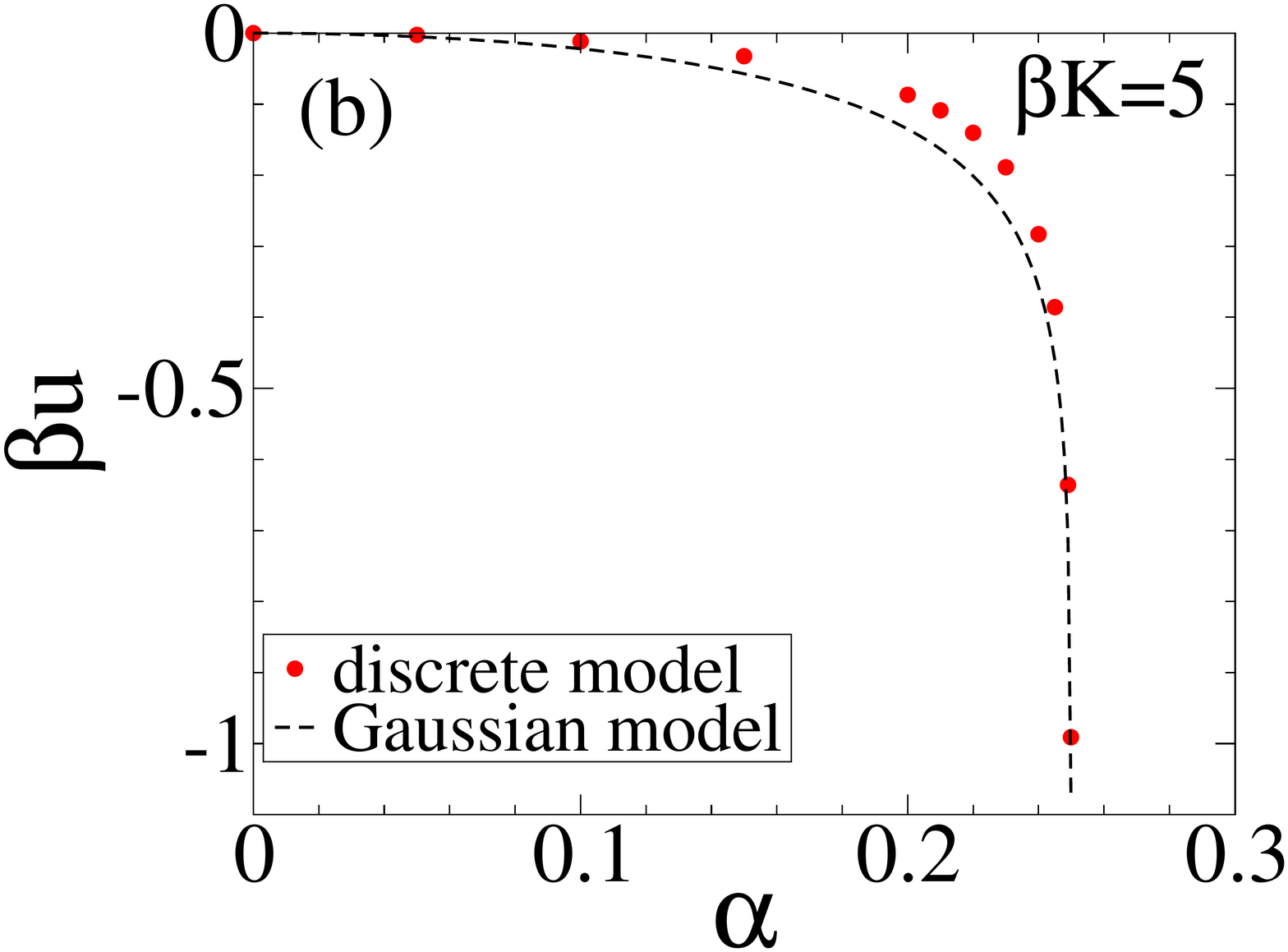}\\
 \end{tabular}
 \end{center}
\caption{The internal energy $u$ as a function of $\alpha$ for 
(a) $\beta K=1$ and (b) $\beta K=5$.   The data points of a discrete Gaussian model are obtained 
from Monte Carlo simulation with $L=128$, and the dashed lines correspond to Eq. (\ref{eq:u}).  }
\label{fig:u12} 
\end{figure}

In Fig. (\ref{fig:u12}) we plot $\beta u$.  The data points are from the Monte Carlo 
simulation for the system 
%of the DG model corresponding to the partition function 
$\Xi_{\infty}$ and the dashed line
%is the same quantity for the continuous Gaussian model according to 
corresponds to the expression in Eq. (\ref{eq:u}).   
For $\beta K=1$, the data points follow closely the continuous Gaussian model.  
For larger $\beta K$, the two results diverge, yet despite this the point of 
thermodynamic instability is the same for both models.  
%that the continuous Gaussian model 
%%%%%%%%%%%%%%%%%%%%%%
\graphicspath{{figures/}}
%%%%%%%%%%%%%%%%%%%%%%%

\begin{figure}[h] 
 \begin{center}
 \begin{tabular}{rrrr}
  \hspace{-0.0cm}    \includegraphics[height=0.20\textwidth,width=0.27\textwidth]{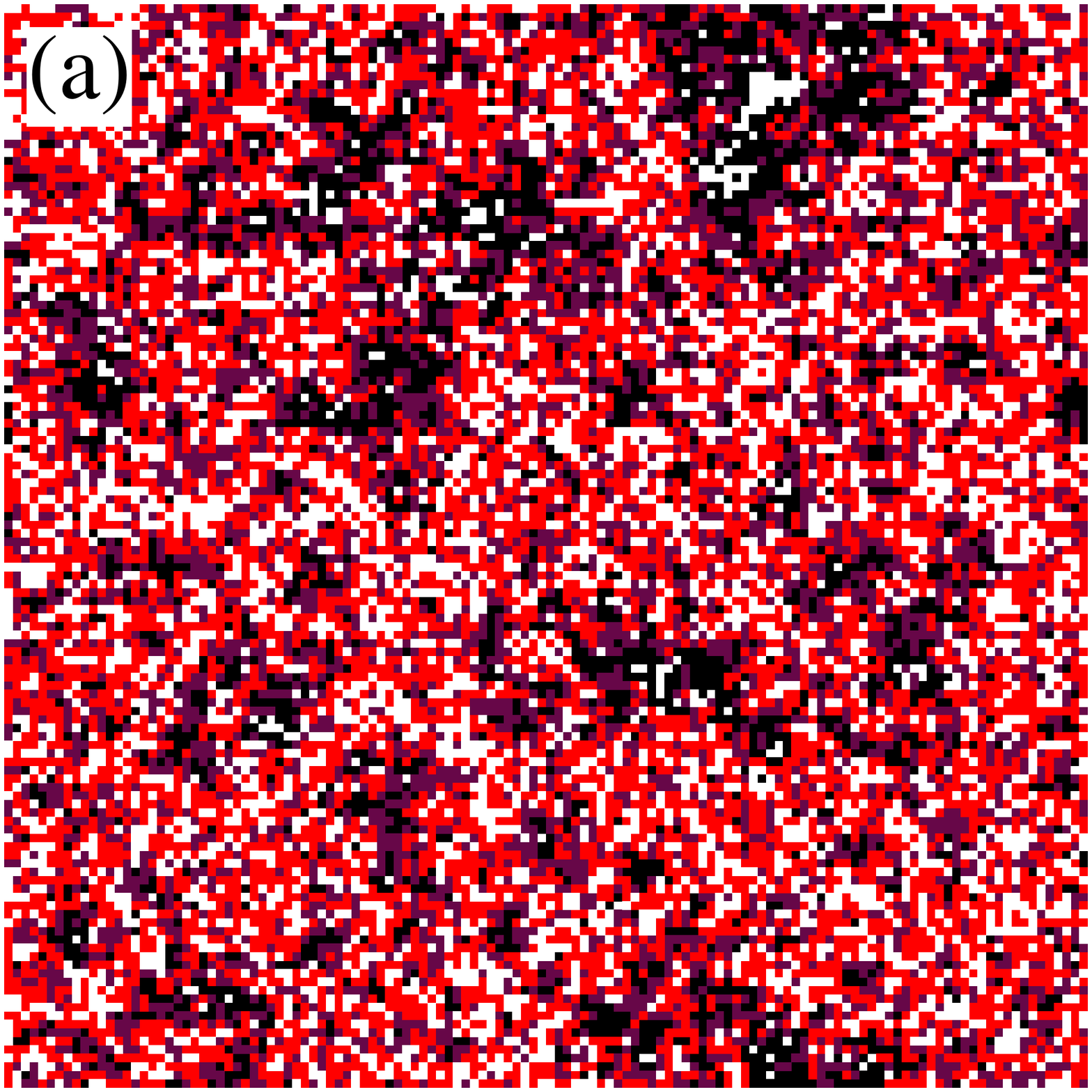}&
  \hspace{-0.5cm}  \includegraphics[height=0.20\textwidth,width=0.27\textwidth]{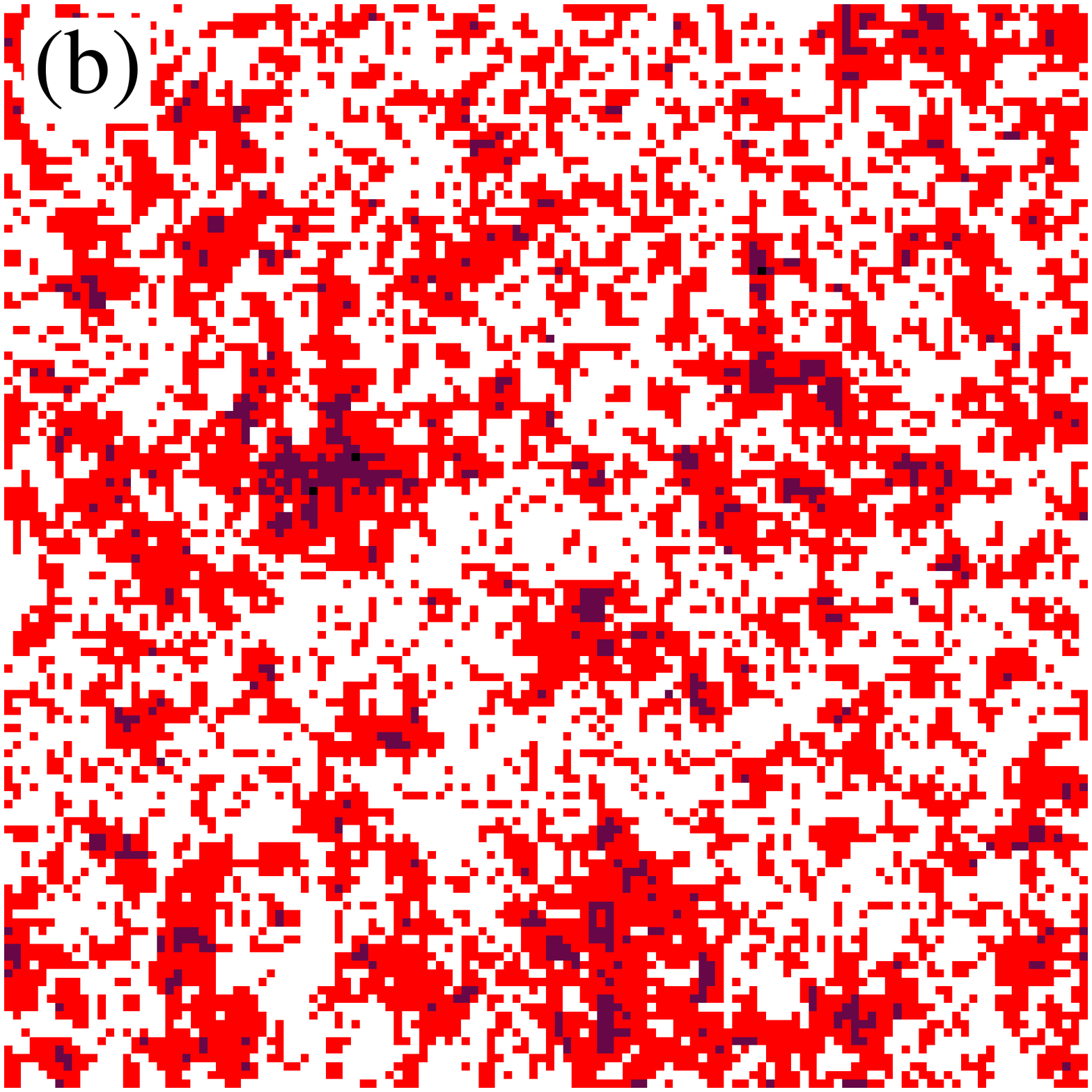}\\
 \end{tabular}
 \end{center}
\caption{Configuration snapshot for $\alpha=0.2499$, for (a) $\beta K=1$ and (b) $\beta K=5$.  
The zero spins are regarded as empty sites and are represented by unfilled squares.  Red is 
for spins $s_i=\pm 1$, purple for $s_i=\pm 2$, and black for $s_i=\pm 3,\pm4,\dots$. }
\label{fig:conf_k1} 
\end{figure}

In Fig. (\ref{fig:conf_k1}) we show configuration snapshots close to thermodynamic 
collapse (at $\alpha=0.2499$) for different values of $K$.  The spin $s_i=0$ is regarded 
as an empty site, and the colored squares are for $s_i\ne 0$.  
%%%%%%%%%%%%%%%%%%%%%%
\graphicspath{{figures/}}
%%%%%%%%%%%%%%%%%%%%%%%

\begin{figure}[h] 
 \begin{center}
 \begin{tabular}{rrrr}
\includegraphics[height=0.20\textwidth,width=0.27\textwidth]{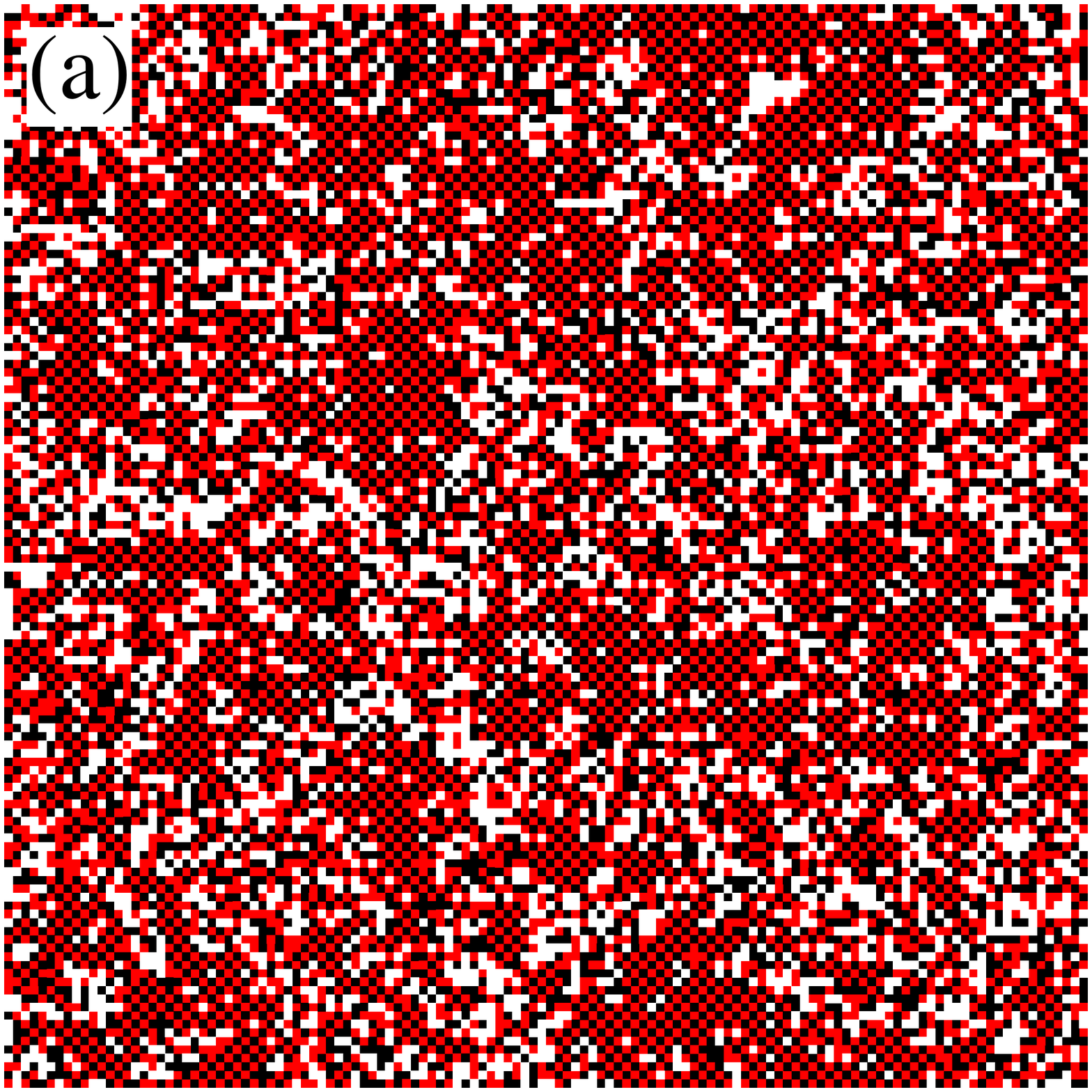}&
  \hspace{-0.5cm}\includegraphics[height=0.20\textwidth,width=0.27\textwidth]{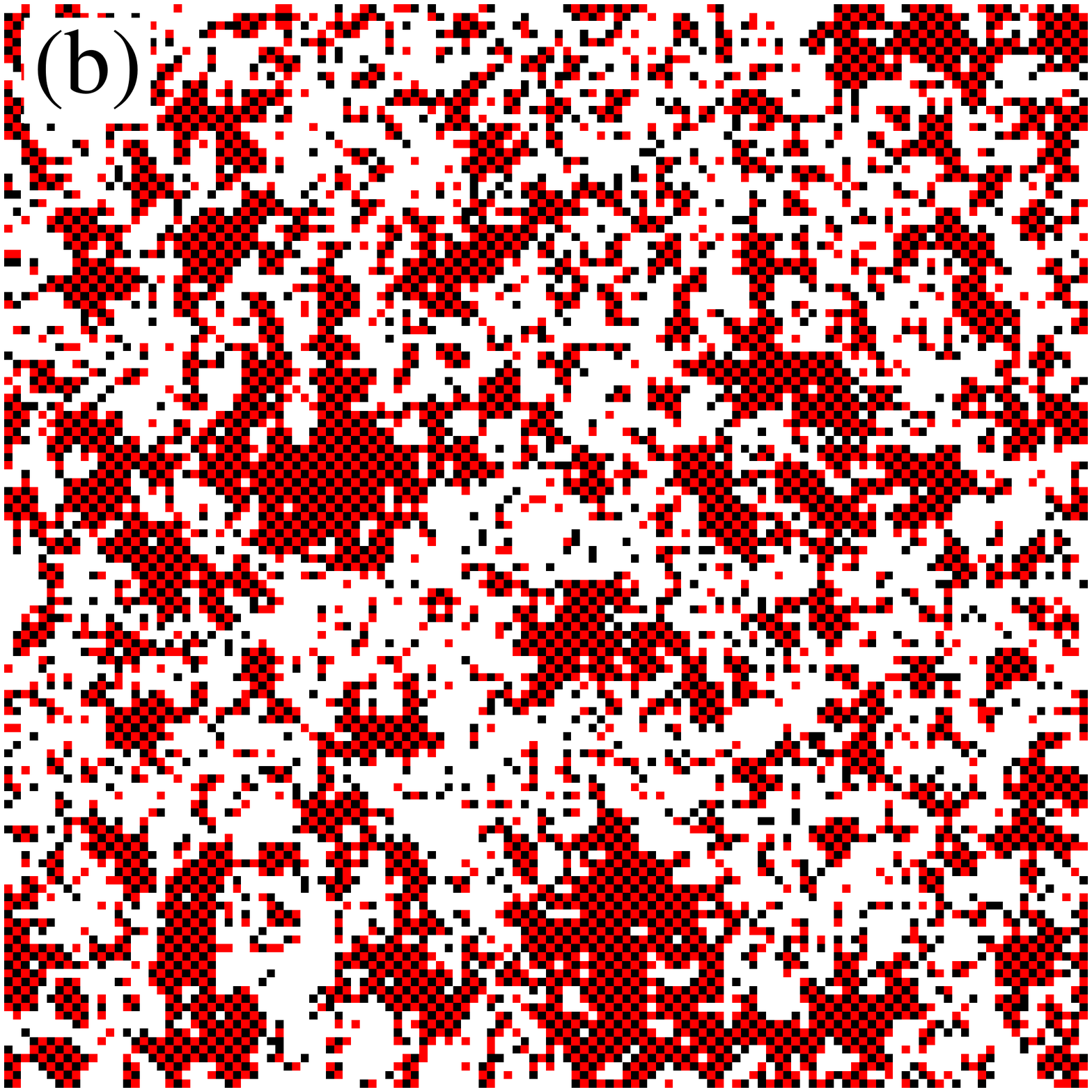}\\
 \end{tabular}
 \end{center}
\caption{Configuration snapshot as in Fig. (\ref{fig:conf_k1}) plotted to 
emphasize ``antiferromagnetic'' order of the configurations.  
The empty sites appear as unfilled squares.  The remaining spins appear as red squares 
if $s_i>0$ and as black squares if $s_i<0$. }
\label{fig:conf_k5} 
\end{figure}

The same configurations are shown in Fig. (\ref{fig:conf_k5}) but in a way as to 
emphasize their antiferromagnetic order.  Red squares are for positive and 
black squares for negative spins.  
%Zero spins are still represented as unfilled squares.  
In both cases, configurations appear as islands of antiferromagnetic material 
immersed in disordered low density phase.  For $\beta K=1$, the islands are much 
larger and appear interconnected, while for $\beta K=5$ the islands
are separated, reminiscent of the liquid-gas coexistence.  
%%%%%%%%%%%%%%%%%%%%%%
\graphicspath{{figures/}}
%%%%%%%%%%%%%%%%%%%%%%%

Another revealing quantity is the distribution of spins at a single site $p(s)$.  For the continuous 
Gaussian model such a distribution is expected to be Gaussian (see Appendix (\ref{sec:A1})
for details), 
\be
p(s) = \frac{e^{-s^2/2\sigma^2}}{\sqrt{2\pi\sigma^2}}.  
\label{eq:pG}
\ee
The variance can be obtained by knowing that the total energy per particle 
for a harmonic system is $\beta u_{tot}=1/2$.  The two contributions to the total energy are 
$\beta u_{tot}=\beta u + \beta u_{ext}$, where $\beta u_{ext} = \int_{-\infty}^{\infty} ds\,p(s) Ks^2/2$ 
and $\beta u$ is given in Eq. (\ref{eq:u}).  This leads to the following result 
\be
\sigma^2 = \frac{2{\rm K}\big(16\alpha^2\big)}{\pi \beta K},  
\label{eq:sigma2}
\ee
and in the limit $\alpha \to \alpha_c$ we have  
\be
\sigma^2 = \langle s^2\rangle \approx -  \frac{1}{\pi \beta K} \ln\bigg(\frac{1-4|\alpha|}{8}\bigg).  
\label{eq:sigma2A}
\ee

In Fig. (\ref{fig:ps}) we plot the distributions $p(s)$ for $\alpha=0.2499$, for different values 
of $\beta K$, and compare the results with the distribution in Eq. (\ref{eq:pG}).  
For $\beta K=1$, the discrete data points coincide with the continuous 
results. 
%while for $\beta K=5$ the data points from the simulation indicate the presence of 
%a larger number of empty sites.  
%%%%%%%%%%%%%%%%%%%%%%
\graphicspath{{figures/}}
\begin{figure}[h] 
 \begin{center}
 \begin{tabular}{rrrr}
\includegraphics[height=0.17\textwidth,width=0.22\textwidth]{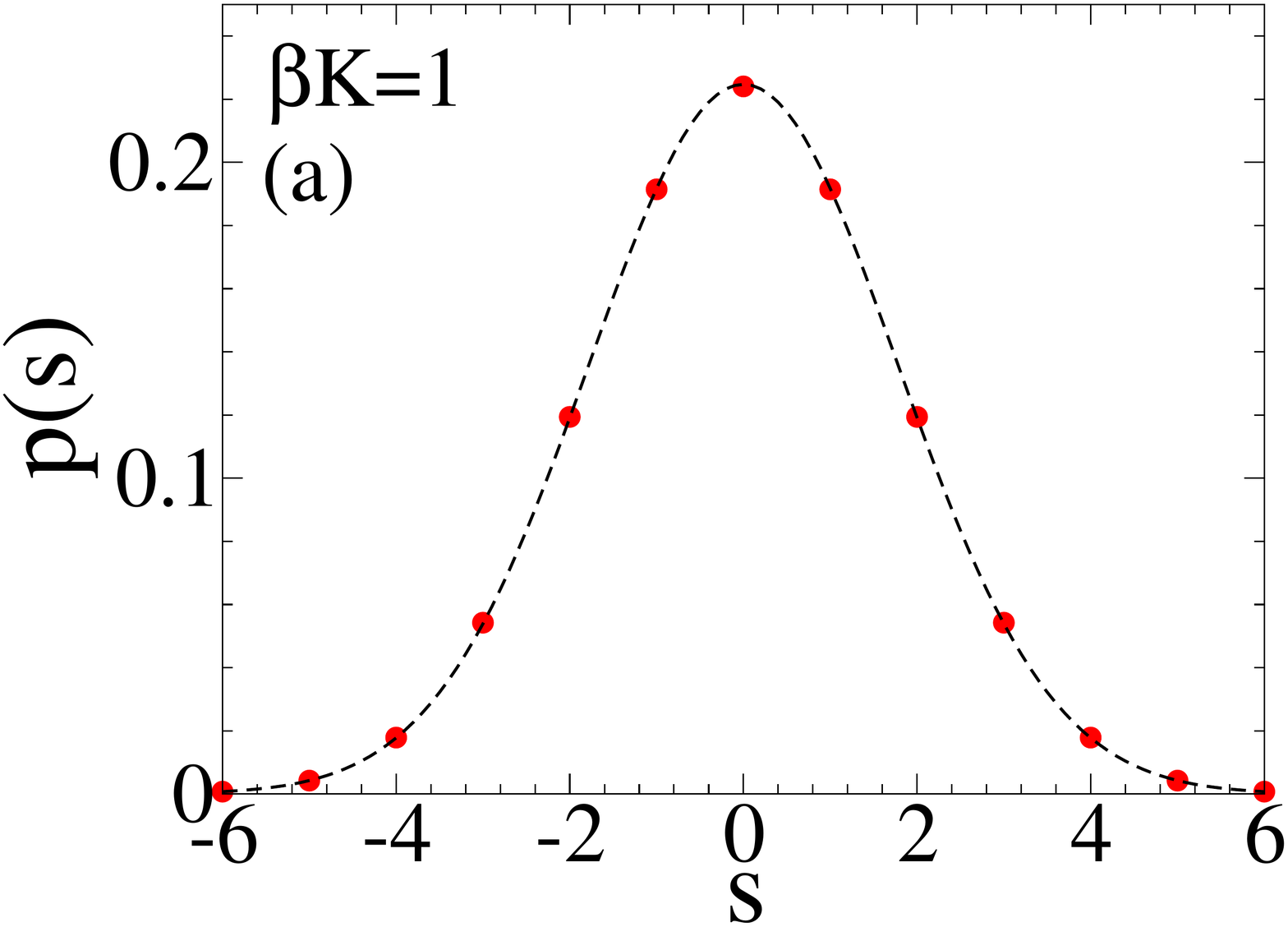}&
\includegraphics[height=0.17\textwidth,width=0.22\textwidth]{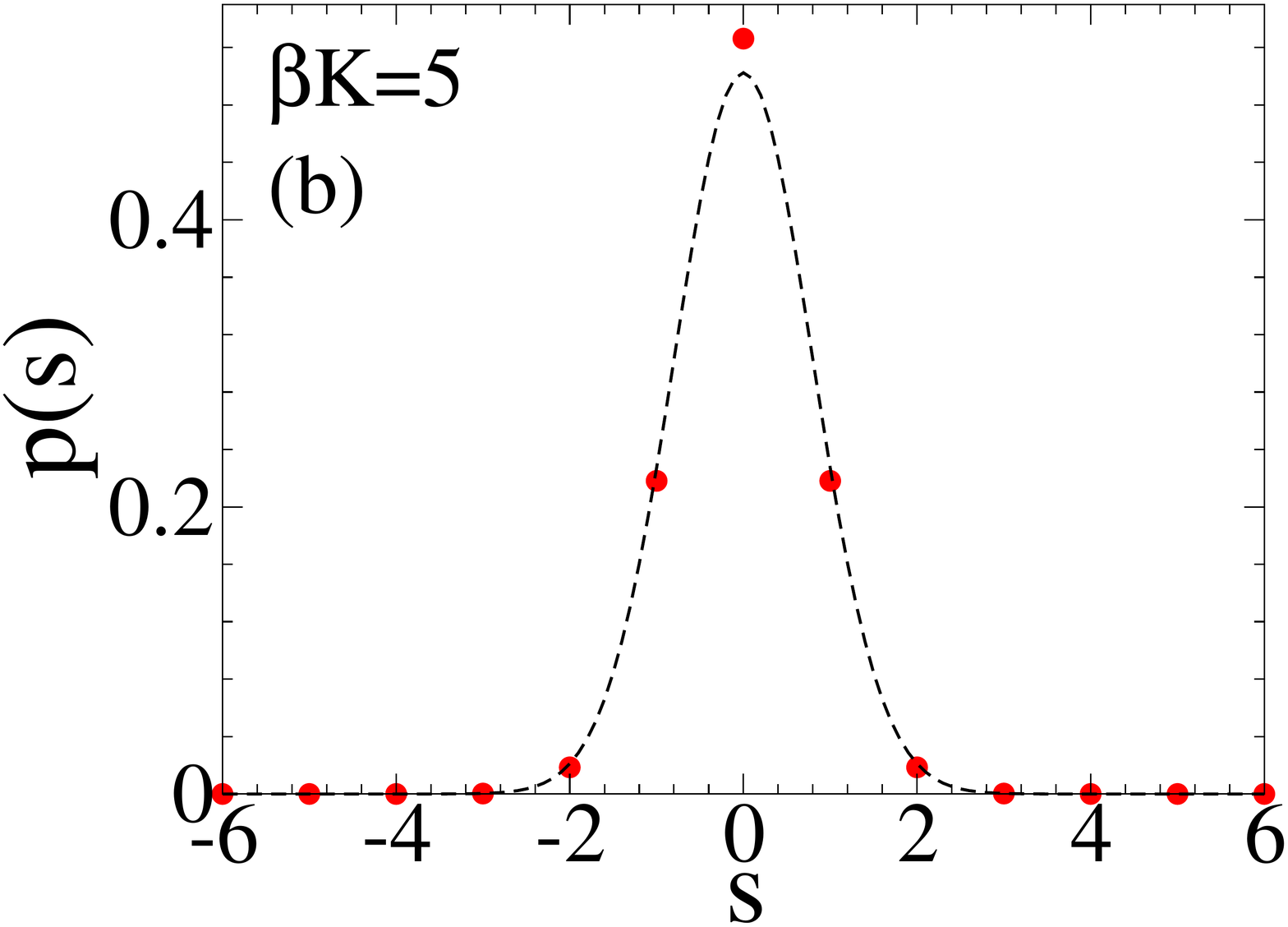}\\
 \end{tabular}
 \end{center}
\caption{Distributions $p(s)$ for $\alpha=0.2499$, and $\beta K=1$ in (a) and $\beta K=5$
in (b).  
The discrete points are from a simulation and the continuous 
lines correspond to Eq. (\ref{eq:pG}). }
\label{fig:ps} 
\end{figure}
%%%%%%%%%%%%%%%%%%%%%%%

%It is reasonable to assume that the distribution 
%of spins at a single site has a Gaussian form, 
%\be
%p(s) = \frac{e^{-s^2/2\sigma^2}}{\sqrt{2\pi\sigma^2}}, 
%\label{eq:pG}
%\ee
%which would mean that the external energy per site is
%\be
%\beta u_{\rm ext} = \frac{K}{2} \int_{-\infty}^{\infty} ds\, p(s) s^2 = \frac{K\sigma^2}{2}.  
%\ee
%From Eq. (\ref{eq:H_ave}) we know that the average total energy per
%site, $u_{\rm tot} = u + u_{\rm ext}$, is $\beta u_{\rm tot}=1/2$.  This  
%allows us to infer the expression for variance as 
%\be
%\sigma^2 = \frac{2{\rm K}\big(16\alpha^2\big)}{\pi \beta K}. 
%\label{eq:sigma2}
%\ee
%Characteristic feature of the continuous Gaussian model is that an average value of a 
%Hamiltonian in Eq. (\ref{eq:H_infty}) is independent of the parameters $K$ and $\alpha$
%and is given by 
%\be
%\langle \beta H_{\infty}\rangle 
%%=K\frac{\partial F_G}{\partial K} 
%= \frac{N}{2}. 
%\label{eq:H_ave}
%\ee
%This is generally not the case for a discrete version of the model.  

\subsection{Discrete subsystem $\Xi_L$}

In the previous section we approximated the $\Xi_{\infty}$ system by neglecting its spin 
discreteness, and the comparison with the simulation showed that such approximation 
is generally correct for $\beta K< 5$, and even if not correct at every point, the CG model 
correctly predicts the point of thermodynamic collapse, suggesting that discreteness has 
no effect on the thermodynamic collapse.  The explanation for this is that close to instability 
the variance of the distribution $p(s)$ diverges, and for large spin variations the spin 
discreteness becomes irrelevant.  

In this section we look more carefully into the neglected contributions of spin 
discreteness by looking into the behavior of $\Xi_L$.  
%It is still interesting to understand how $\Xi_L$ contributes to an overall system.  
%Based on Fig. (\ref{fig:u12}), these contributions increase with increasing $\beta K$ 
%(or the effective temperature according to Eq. (\ref{eq:Xi_L})).  
%Furthermore, 
%The reason why this is interesting is that 
According to Ref. \cite{Chui76}, the DG model at $\alpha=\alpha_c$ is isomorphic 
with the lattice Coulomb model which exhibits the Kosterlitz-Thouless (KT) transition.  
This means that at precisely the point where our system is about to collapse, the
system also undergoes the KT transition along the parameter $\beta K$ \cite{Gupta97}.  
This by itself cannot affect the collapse transition, however, it can modify the manner 
of that collapse.

\subsubsection{$\Xi_L$ in one-dimension}
To establish the procedure in a clear manner, 
we consider first a simpler case of a system in $d=1$, for which the matrix $A$ is given in 
Eq. (\ref{eq:A1}) and the matrix $A^{-1}$ is 
\be
A^{-1}_{ij}  = \frac{1}{L} \sum_{k=0}^{L-1} \frac{\cos\big[2\pi k (i-j) / L\big]}{1+2\alpha\cos(2\pi k/L)}.  
\label{eq:AI_1d}
\ee
Because the value of $\alpha_c$ depends on dimensionality according to $\alpha_c = 1/(2d)$, 
in $d=1$ thermodynamic collapse occurs for $\alpha_c=1/2$.  

In the limit $L\to\infty$ the summation in Eq. (\ref{eq:AI_1d}) becomes an integral,  
\be
A^{-1}_{ij}  = \frac{1}{2\pi} \int_0^{2\pi} dq\, \frac{\cos\big[q(i-j)\big]}{1+2\alpha\cos(q)}, 
\label{eq:AD1}
\ee
which evaluates to
%$A^{-1}$, which plays the role of the interaction 
%potential in $\Xi_L$, is given by 
\be
A^{-1}_{ij}   = \frac{(-1)^{|i-j|}} {\sqrt{1-4\alpha^2}} \bigg(\frac{1-\sqrt{1-4\alpha^2}}{2\alpha}\bigg)^{|i-j|}.  
\label{eq:AD2}
\ee
%$A^{-1}$, which plays the role of an interaction potential between two sites at separation $|i-j|$,  
%decays exponentially with $|i-j|$ if $\alpha<\alpha_c=1/2$.  
At $\alpha_c$, $A_{ij}^{-1}$ diverges, but the divergence can be subtracted and the system 
can be analyzed in terms of non-divergent interactions.  
To do this, we introduce an alternating sign matrix, 
\be
C^{}_{ij}  =  (-1)^{|i-j|},
\ee
then subtract from each element $A_{ij}^{-1}$ the divergent term $C_{ij}/\sqrt{1-4\alpha^2}$.  
%\be
%C^{}_{ij}  = \bigg(\frac{1}{\sqrt{1-4\alpha^2}}\bigg) (-1)^{|i-j|}. 
%\ee
The remaining elements constitute an interaction matrix 
$U_{ij}  = A_{ij}^{-1} - C_{ij}/\sqrt{1-4\alpha^2}$, 
%\be
%U_{ij}   = \frac{1}{\sqrt{1-4\alpha^2}} \bigg[ \bigg(\frac{-1 + \sqrt{1-4\alpha^2}}{2\alpha}\bigg)^{|i-j|} - (-1)^{|i-j|}\bigg],
%\ee
which at $\alpha=\alpha_c$ reduces to 
\be
U^{}_{ij}  =  -(-1)^{|i-j|}   |i-j|.  
\ee
%which, if we neglect an alternating sign, is a Coulomb interaction in 1D.  

The Hamiltonian of the system $\Xi_L$ can now be written as 
\be
\beta H_L = \frac{2\pi^2}{\beta K} \bigg[{s}^T U {s} + \frac{{s}^T C {s}}{\sqrt{1-4\alpha^2}}\bigg].  
\ee
Clearly, only configurations which suppress the divergence are allowed.  
Such configurations satisfy ${s}^T C=0$, which is the same as 
%yield ${s}^T C_{}{s}=0$, or such that the elements 
%of a vector $s^TC$ are $[s^TC]_{i} = 0$, 
%\be
%\sum_{j=1}^{N} (-1)^{|i-j|} s_j = 0
%\ee
\be
\sum_{odd} s_i = \sum_{even} s_i,
\label{eq:res}  
\ee
where the subscripts ``odd'' and ``even'' refer to odd and even numbered lattice sites.  
Taking this restriction into account, the Hamiltonian can now be written as  
\be
\beta H'_L = \frac{2\pi^2}{\beta K} {s}^T U {s}, 
\ee
where the prime implies the restriction in Eq. (\ref{eq:res}).  
%The question is, is there anything interesting behavior as the 
%effective temperature $k_BT_{eff} = \beta K/(4\pi^2)$ varies.  

Although not immediately clear, $\Xi_L$ is an even function of $\alpha$, and 
flipping the sign of $\alpha$ does not change the partition function.  
(The sign change modifies Eq. (\ref{eq:AD2}), but as the summations in $\Xi_L$
are over $s_i\in(-\infty,\infty)$, this does not effect the value of $\Xi_L$).  
Calculations then can equally be done for $\alpha=-1/2$.  In such a case, 
the interaction potential becomes 
\be
U^{}_{ij}  =  -|i-j|,
\ee
which is a Coulomb interaction in 1D.  There are two differences between the present 
system and the more usual Coulomb model, however.   First, the valance number of 
particles on a lattice site is unlimited.  Second, the periodic boundary conditions 
involve only particles in the simulation box and do not include contributions due to images 
outside the original simulation box.

\subsubsection{$\Xi_L$ in two-dimensions}
Based on the results of the previous section for $d=1$, 
it is guessed that in $d=2$ the interactions between 
lattice sites are logarithmic at $\alpha_c$, since this is the functional form of 
Coulomb interactions in this dimension.  
It is more convenient to represent interactions between spins on a square-lattice, not 
in terms of the matrix $A^{-1}$, but in terms of a pair potential between sites on the
$(x,y)$-grid, and such a potential would have the following form \cite{Chui76} 
\be
U_{tot}  = \bigg(\frac{1}{2\pi}\bigg)^2 \!\! \int_{-\pi}^{\pi} \!\! dq_1\int_{-\pi}^{\pi} \!\! dq_2\, \frac{\cos(q_1n + q_2m) }{1 + 2\alpha\cos(q_1) + 2\alpha\cos(q_2)},  
\ee
where $n=|x_1-x_2|$ and $m=|y_1-y_2|$ indicate a separation between two 
lattice sites on the discrete Cartesian grid, where $x_i,y_i=0,1,\dots$.  
The expression is analogous to that in Eq. (\ref{eq:AD1}) for $d=1$ in the limit $L\to\infty$.

If we expand the integrand in powers of $\alpha$ and then evaluate each term, we find the following 
series expansion 
\be
U_{tot}(m,n) = \sum_{k=0}^{\infty} \frac{ (-1)^{m+n} \alpha^{2k+m+n} (2k+m+n)!^2}{k!(k+m)!(k+n)!(k+m+n)!}, 
\label{eq:U_tot}
\ee
which constitutes a hypergeometric function.  $U_{tot}(m,n)$ diverges at $\alpha_c=1/4$, and the 
divergent term is identified as 
\be
U_{tot}(0,0) = \frac{2{\rm K} \big(16\alpha^2\big)}{\pi},
\label{eq:A00}
\ee
where ${\rm K}(x)$ is the complete elliptic integral of the first kind.  Subtracting the divergence from 
$U_{tot}$, the non-divergent pair potential is 
\be
U(m,n) = U_{tot}(m,n) - (-1)^{m+n}\frac{2{\rm K} \big(16\alpha^2\big)}{\pi}, 
\label{eq:U_2D}
\ee  
where an accurate approximation to $U(m,n)$ at $\alpha_c$ is \cite{Spitzer}
\be
U(m,n) \approx -(-1)^{m+n} \frac{2}{\pi} \bigg(\ln\sqrt{n^2 + m^2} + \gamma + \frac{1}{2}\ln 8\bigg), 
\label{eq:U_2D_approx}
\ee
that is valid for $\sqrt{n^2 + m^2}\ge 1$.  For $n=m=0$ we use $U=0$.  
The approximate functional form in Eq. (\ref{eq:U_2D_approx}) compared with the exact
form in Eq. (\ref{eq:U_2D}) is shown in Fig. (\ref{fig:ps}).
%%%%%%%%%%%%%%%%%%%%%%
\graphicspath{{figures/}}
\begin{figure}[h] 
 \begin{center}
 \begin{tabular}{rrrr}
\includegraphics[height=0.17\textwidth,width=0.22\textwidth]{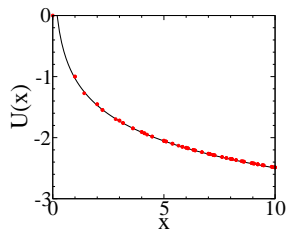}&
 \end{tabular}
 \end{center}
\caption{Approximate pair potential in Eq. (\ref{eq:U_2D_approx}) compared to the
exact results in Eq. (\ref{eq:U_2D}) at discrete locations on a square-lattice. }
\label{fig:ps} 
\end{figure}
%%%%%%%%%%%%%%%%%%%%%%%

Because the constant terms in $U(m,n)$, together with the divergent term, are irrelevant, 
the pair interaction can simply be written as 
\be
U = -(-1)^{m+n} \frac{2}{\pi} \ln\sqrt{n^2 + m^2}.  
\ee
The spin configurations are subject to the same 
restriction as that in Eq. (\ref{eq:res}).  In the square-lattice setting, this means that the 
lattice is decomposed into two interpenetrating sub-lattices and the restriction amounts to 
$\sum_{sub_1} s_i =  \sum_{sub_2} s_i$.

The Hamiltonian at $\alpha_c$ can be written as 
\ba
\beta H'_L &=& -\frac{4\pi}{\beta K} \sum_{x_1,y_1=1}^{L}  \sum_{x_2,y_2=1}^{L}(-1)^{m+n}  
s_{x_1,y_1}s_{x_2,y_2} \nonumber\\
&\times& \ln\sqrt{n^2 + m^2}, 
%\beta H'_L = -\frac{4\pi}{\beta K} \sum_{i}^{L^2}  \sum_{j}^{L^2} (-1)^{m+n}  s_{i} s_{j} \ln\sqrt{r_{ij}}, 
\label{eq:HL_2D}
\ea
with $x_i$ and $y_i$ indicating discrete locations on a lattice grid.

\begin{figure}[h] 
 \begin{center}
 \begin{tabular}{rrrr}
\includegraphics[height=0.20\textwidth,width=0.26\textwidth]{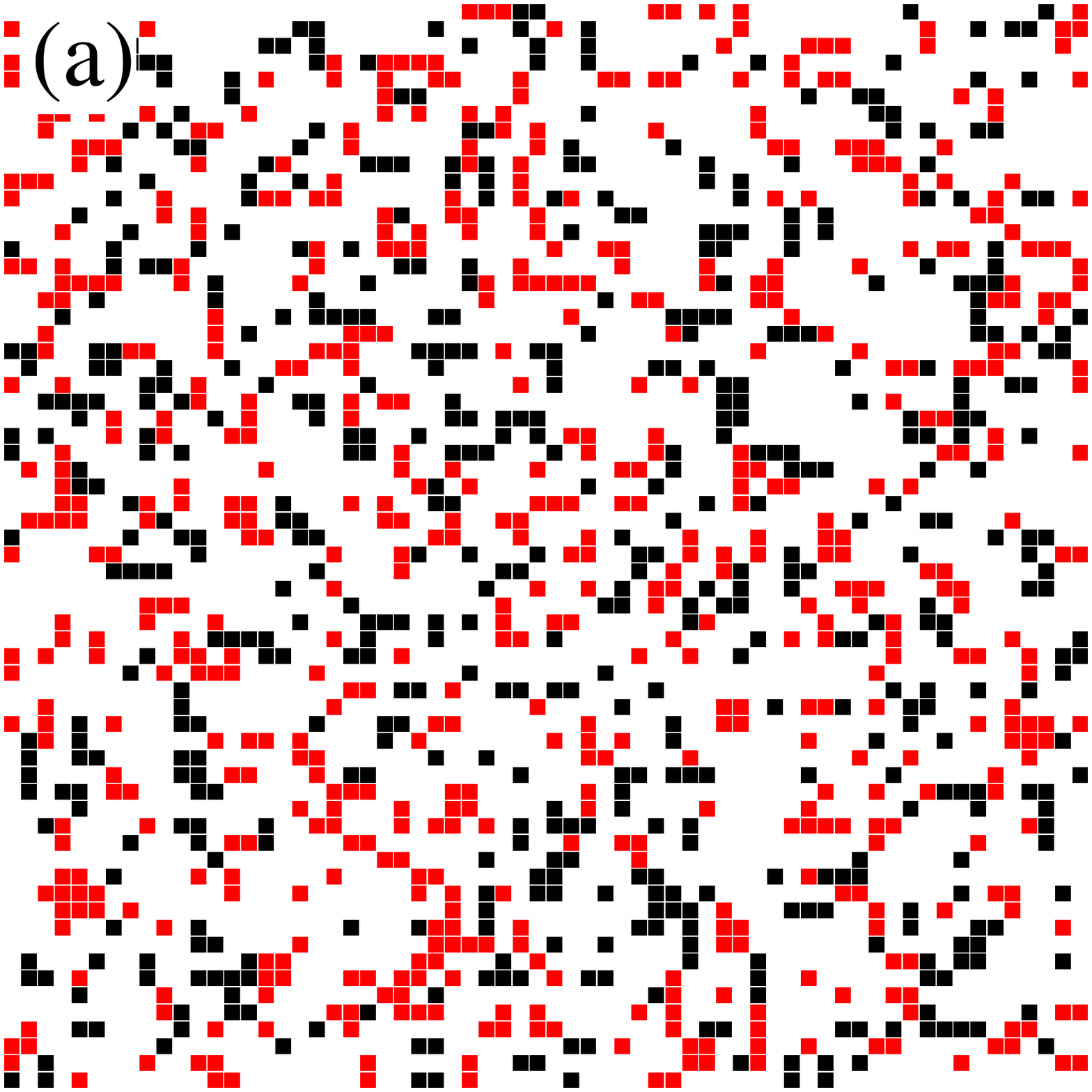}&
\includegraphics[height=0.20\textwidth,width=0.26\textwidth]{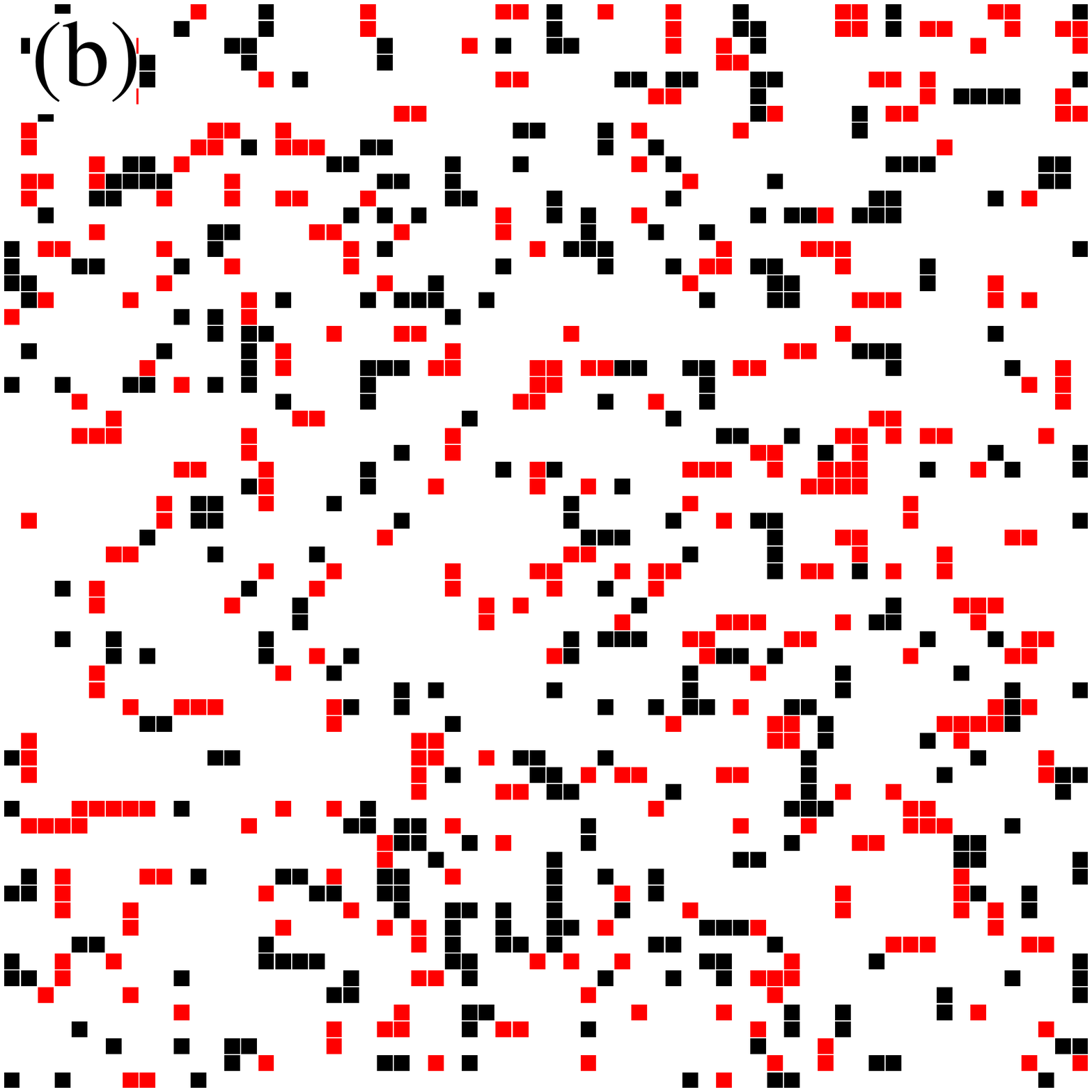}\\
\includegraphics[height=0.20\textwidth,width=0.26\textwidth]{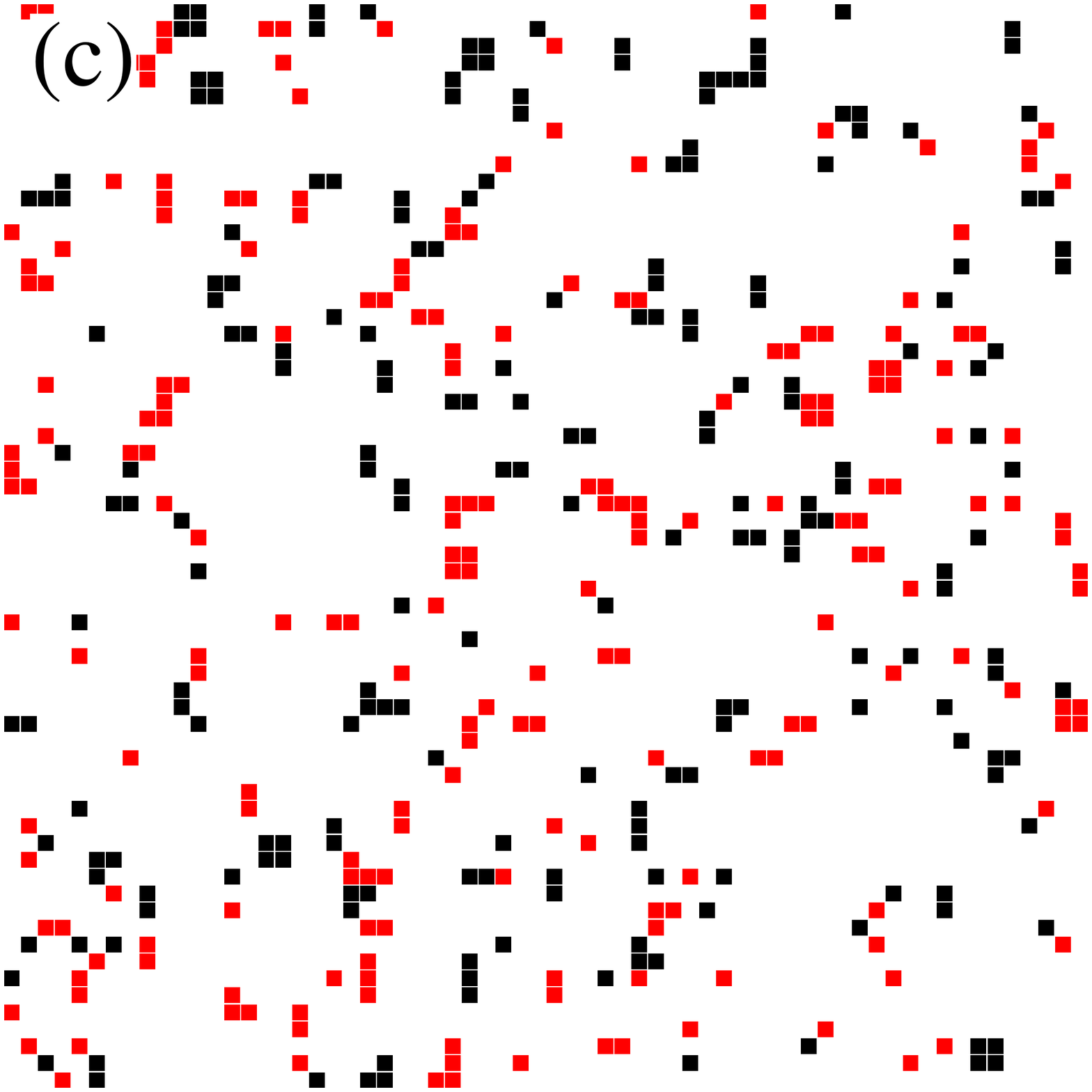}&
\includegraphics[height=0.20\textwidth,width=0.26\textwidth]{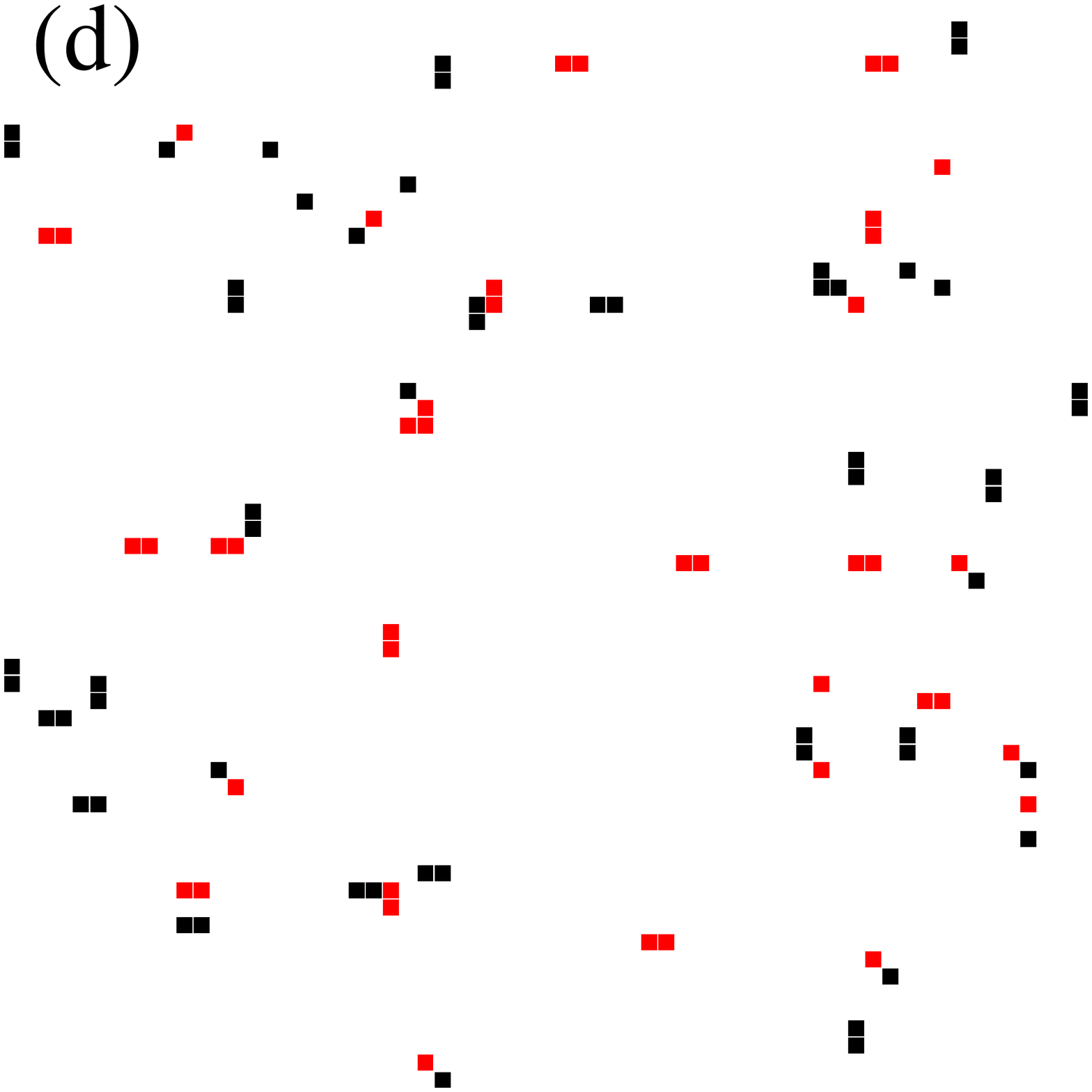}\\
 \end{tabular}
 \end{center}
\caption{Configuration snapshots for $\Xi_L$ at $\alpha_c$ for different values of 
$\beta K$, $\beta K=12,10,8,6$.  Red squares are for $s=1$, black squares for $s=-1$,
and the white squares represent empty sites.  Spins larger than $1$ are 
negligible for those values of $\beta K$ that are plotted.  
The system size is $L=64$.}
\label{fig:confL} 
\end{figure}

In Fig. (\ref{fig:confL}) we show several configuration snapshots of $\Xi_L$ for decreasing 
values of $\beta K$.  One observes gradual decrease of spin 
density with decreasing $\beta K$, and for $\beta K=6$ the configuration consists of 
sparse isolated spins or spin pairs of the same sign.  This means that for $\beta K<6$, 
$\Xi_L\approx 1$ since most likely value of a spin is $s_i=0$.  
%%%%%%%%%%%%%%%%%%%%%%%
\graphicspath{{figures/}}
%%%%%%%%%%%%%%%%%%%%%%%

\begin{figure}[h] 
 \begin{center}
 \begin{tabular}{rrrr}
\includegraphics[height=0.17\textwidth,width=0.22\textwidth]{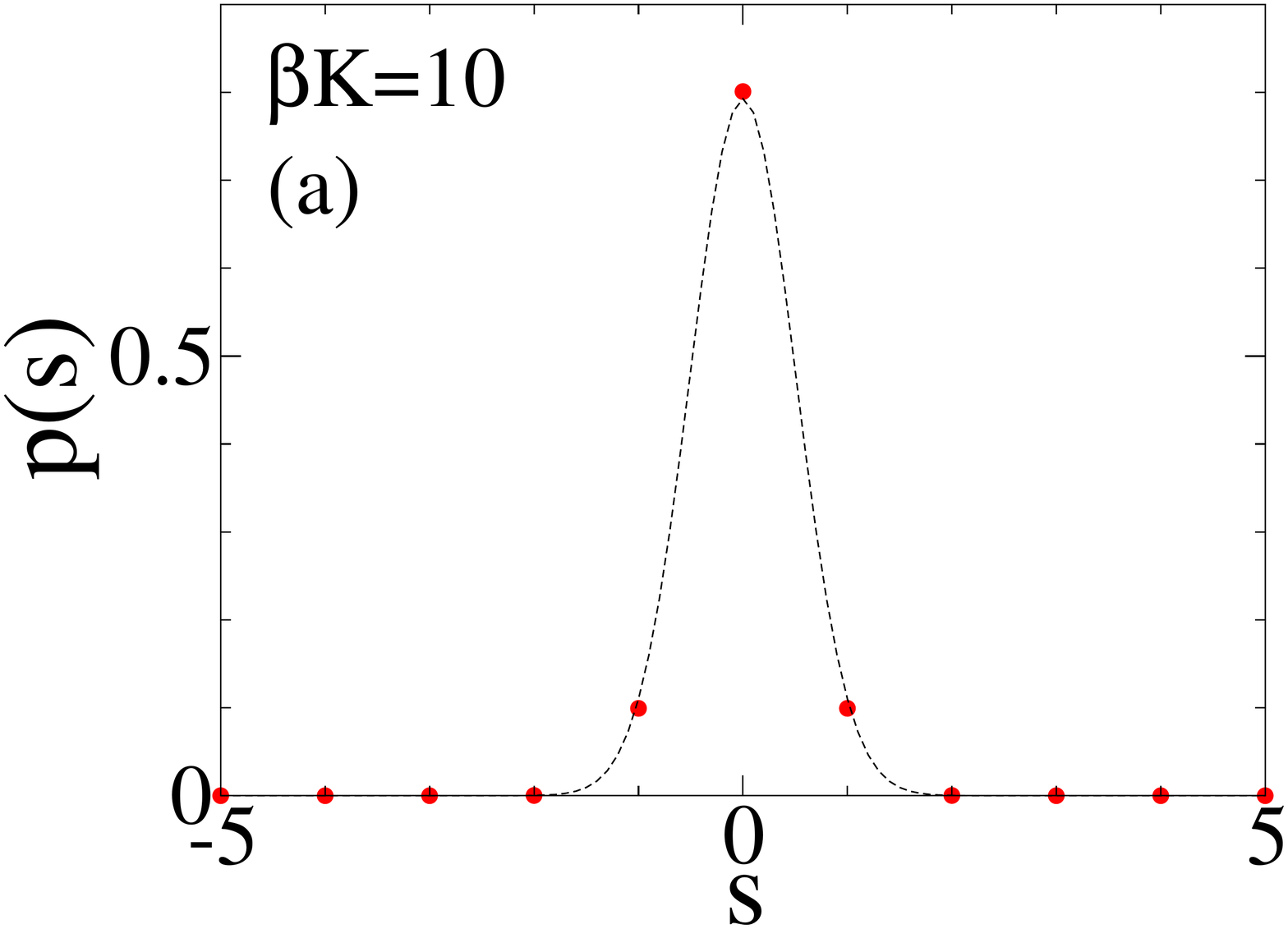}&
\includegraphics[height=0.17\textwidth,width=0.22\textwidth]{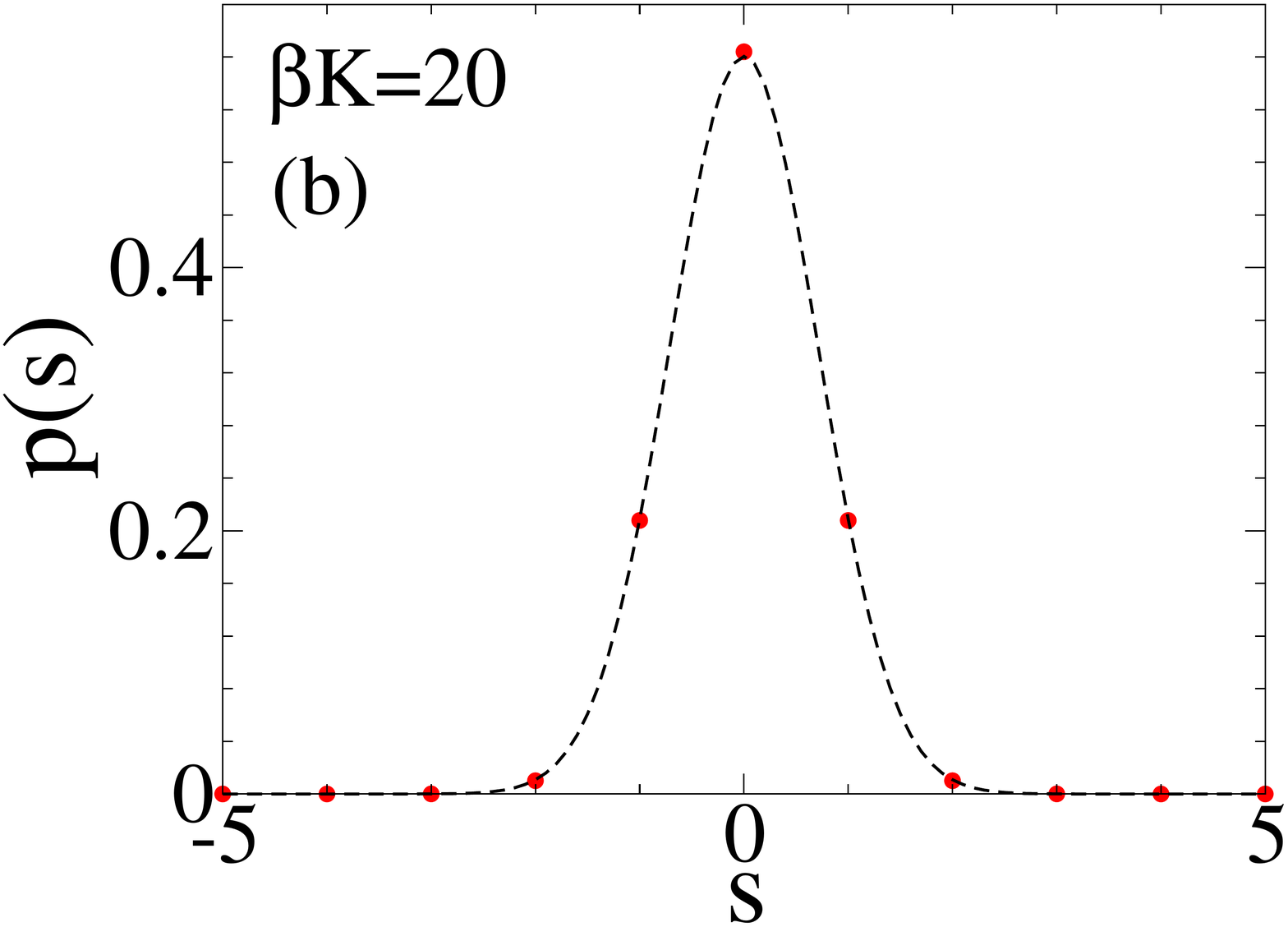}\\
 \end{tabular}
 \end{center}
\caption{Distributions $p(s)$ for a system $\Xi_L$ at $\alpha=\alpha_c$, for (a)
$\beta K=10$ and (b) $\beta K=20$.  
The discrete points are from a simulation and the continuous 
lines correspond to Eq. (\ref{eq:pL}). }
\label{fig:psL} 
\end{figure}

The distribution of spins $p(s)$ is accurately represented 
using the continuous Gaussian approximation, see Appendix (\ref{sec:A0}), given by 
\be
p(s) = e^{-2\pi^2 s^2/\beta K} \sqrt{\frac{2\pi}{\beta K}},
\label{eq:pL}
\ee
where the spin variance is given by $\langle s^2\rangle = \frac{\beta K}{4\pi^2}$.  
Fig. (\ref{fig:psL}) compares the above Gaussian distribution with the discrete distributions 
obtained from simulation, showing a general good agreement between the two.  

The Gaussian distribution in Eq. (\ref{eq:pL}), however, cannot be a reliable approximation 
of the discrete system if 
%a continuous distribution, but when 
%applied to discrete spins, it does not make sense if 
$p(0)>1$, since this implies that the probability that a spin is zero is greater than one.   
%or, since 
%\be
%p(0)=\sqrt{\frac{2\pi}{\beta K}},
%\label{eq:p0L}
%\ee
The Gaussian approximation in Eq. (\ref{eq:pL}), therefore, breaks down for $\beta K < 2\pi$.   
%On the other hand, 
%The approximation is highly accurate for $\beta K>2\pi$
%as seen in Fig. (\ref{fig:psL}), which plots $p(s)$ obtained in a simulation and compares 
%them with Eq. (\ref{eq:pL}).
%%%%%%%%%%%%%%%%%%%%%%
\graphicspath{{figures/}}
%%%%%%%%%%%%%%%%%%%%%%%

\begin{figure}[h] 
 \begin{center}
 \begin{tabular}{rrrr}
\includegraphics[height=0.17\textwidth,width=0.22\textwidth]{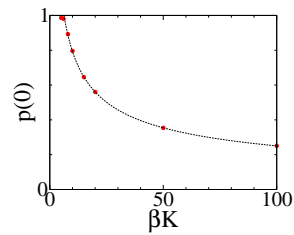}\\
 \end{tabular}
 \end{center}
\caption{The probability that a lattice site is empty, $p(0)$, as a function of $\beta K$.  
The data points are from simulation and the continuous line corresponds to $p(0)$
in Eq. (\ref{eq:pL}). }
\label{fig:p0L} 
\end{figure}

In Fig. (\ref{fig:p0L}) we plot $p(0)$ as a function of $\beta K$ obtained from 
simulation for a discrete system and compare it to $p(0)$ calculated using Eq. (\ref{eq:pL}).  
%%%%%%%%%%%%%%%%%%%%%%
\graphicspath{{figures/}}
%%%%%%%%%%%%%%%%%%%%%%%

Given the reliable performance of the approximation in the range $\beta K>2 \pi$, it is 
safe to conclude that there is no phase transition in this range.  The distribution 
$p(s)$ is monomodal and its variance diverges only in the limit $\beta K\to \infty$.   
If there is any KT type of transition, it must occur in the range $5 > \beta K > 6$ 
and can be associated with the emergence of the lone pairs in Fig. (\ref{fig:confL}), 
which could be interpreted as the emergence of defects.  

Because the MC simulations on the spin ensemble become impossible for 
$\beta K>5$, since the only possible spins are $s_i=0$, the KT transition 
along the $\beta K$ parameter in the context of the two-component model could imply a 
different mechanism of the collapse transition.

%and the same problem emerges 
%for the MC simulations for the Coulomb lattice system for $\beta K<6$

%\newpage
\section{Finite $\rho$ and the emergence of a metastable region}
\label{sec:rho}

In this section we consider a more realistic situation where the average occupation 
number of a lattice site $\rho$ is finite.  This also means that the quadratic Hamiltonian 
$H_{\infty}$ in Eq. (\ref{eq:H_infty}) is modified by an additional non-quadratic term $h(s)$.  
A technical difficulty is that the system is no longer Gaussian and additional methods are 
needed to analyze it.  

The simulations show that the thermodynamic collapse for finite
$\rho$ does not occur at $\alpha=1/4$, as for the case $\rho\to\infty$,   
but is shifted to larger values of $\alpha$.  
This indicates that the thermodynamic collapse depends on density.  
This may be somewhat surprising, since one expects the global minimum of a system 
for $\alpha>1/4$ to be a collapsed state.  This indicates the presence of a metastable
equilibrium.  

%A physical consequence is that the region of stability starts to shift 
%with decreasing density;  simulations in both canonical and grand canonical ensemble 
%show that the system does not spontaneously collapse after crossing $\alpha=1/4$, but 
%at another point $\alpha>1/4$ that depends on density.  

In a two-component system, a collapsed configuration, as it emerges in a simulation, is comprised of 
numerous clusters, each of which can, in principle, accommodate an infinite number of particles. 
A sequence of such clusters for a one-dimensional lattice model has been analyzed before 
\cite{Frydel18b}.  Within a single cluster, a single site is occupied by one type of particles.  (Similar 
clusters have been observed in a two-component system of penetrable spheres \cite{Frydel17,Frydel18a}).  
The energy of each cluster scales like $E \propto - n^2$, where $n$ is the number 
of particles in a cluster.  

If a collapsed configuration consists of a single cluster comprised
of all the particles in a system, then the energy of a collapsed state scales like 
$E \propto - n^2$ where $n$ is the 
number of particles.  The competing entropy of non-collapsed configurations, on 
the other hand, 
scales like $-ST \propto k_BT n\ln n$.  This means that as soon as a configuration with 
energy that scales like $E \propto - n^2$ appears (which for the present model occurs when 
$\alpha>1/4$),  the global minimum will always be a collapsed state.  The fact that the system 
does not collapse spontaneously when $\alpha>1/4$ suggests 
that there is a local minimum that produces metastable equilibrium.

For a better grasp of the collapse mechanism, we describe a simple situation.  We consider 
a finite system that roughly corresponds to a size of a cluster that emerges in a collapsed state. 
The system is in contact with a reservoir, so that a number of particles in the system $n$ fluctuates.  
The particles in the reservoir do not interact with each other, while 
the energy of the system itself is assumed to be $\beta E=-an^2$ so that the system can 
achieve a collapsed configuration only if $a>0$.  %Assuming a binary species, 
%The cluster is in contact with a reservoir and 
The grand potential of the system is 
\be
\beta \Omega(n) = n\ln\frac{n}{2} - n - a n^2 - \beta\mu n, 
\label{eq:omega}
\ee
where $n=n_+ + n_-$ is the total number of particles and $\frac{n}{2}\ln\frac{n}{2}-\frac{n}{2}$ is the entropy 
$-TS$ due to each species.  
If $a>0$, the global minimum of $\beta \Omega(n)$ is for $n=\infty$.
However, there is also a local minimum $\frac{d\beta\Omega}{dn}=0$ corresponding to  
\be
n_0 = -\frac{W\big(-4 a e^{\beta\mu}\big)}{2a} =  2 e^{\beta\mu} + 8a  e^{2\beta\mu} + \dots
\ee
and that corresponds to a metastable equilibrium.  The local minimum vanishes for 
$a > \frac{1}{4e^{\beta \mu}}$.  
%The extent of the metastable region increases with smaller density of a reservoir and can become arbitrarily 
%large for a very dilute system. 
Since the reservoir density is given by $\rho=e^{\beta \mu}$, then the thermodynamic collapse 
can be estimated to depend on the density as $a = \frac{1}{4\rho}$.  We observe a similar 
qualitative behavior in our simulations for a lattice-gas model of binary penetrable particles.   
%Similar general behavior is observed in the original lattice system as will be shown below.  

To use a more rigorous approach to analyze a metastable region, we start with a perturbation 
approximation.  For a finite $\rho$, the system Hamiltonian is 
\be
H =  H_{\infty} +  \sum h(s_i).
\ee
The partition function of this system can be written in terms of the $H_{\infty}$ ensemble as 
\cite{Frydel15}
\be
\Xi = \Xi_{\infty}\big\langle e^{-\beta \sum h(s_i)}\big\rangle_{\infty}.  
\ee
If we expand the quantity $\ln \Xi$, assuming $h$ to be small, and keep only the first order term, 
a perturbative expression is 
\be
\ln \Xi \approx \ln \Xi_{\infty} - \beta N \big\langle h(s)\big\rangle_{\infty}.  
\ee
Finally, if we use the separation $\Xi_{\infty}=\Xi_{G}\Xi_{L}$ and ignore discrete contributions, 
$\Xi_{\infty}\approx \Xi_{G}$, we have 
\be
\ln \Xi \approx \ln \Xi_{G} - \beta N \big\langle h(s)\big\rangle_{G},
\ee
where the subscript $G$ denotes the continuous Gaussian system analyzed earlier.

For indistinguishable particles $h(s_i) = -\mu' |s_i|$, where 
the average value of $|s_i|$ is related to $A^{-1}_{ii}$, see Eq. (\ref{eq:A3_3}), 
and the value of $A^{-1}_{ii}$ is given in Eq. (\ref{eq:A00}) for $A^{-1}_{ii}$, we get 
\be
\ln \Xi \approx  \ln \Xi_G + \beta \mu' \frac{2N}{\pi}  \sqrt{\frac{{\rm K} (16\alpha^2)}{\beta K}}.   
\label{eq:Xi_p1}
\ee
The internal energy per particle can now be obtained using the definition in Eq. (\ref{eq:u0}).  
For $\beta \mu'<0$, the expression in Eq. (\ref{eq:u})  
is corrected as $u\to u + \Delta u$, where the correction due to the perturbation theory is given by 
\be
\beta \Delta u = \frac{\beta \mu' }{\pi}  \sqrt{\frac{{\rm K} (16\alpha^2)}{\beta K} }
\bigg[ 1 - \frac{1}{(1-16\alpha^2)} \frac{{\rm E} (16\alpha^2)}{{\rm K} (16\alpha^2)} \bigg],
\label{eq:du}
\ee 
where ${\rm E}(x)$ is the complete elliptic integral of the second kind.  

\begin{figure}[h] 
 \begin{center}
 \begin{tabular}{rrrr}
  \includegraphics[height=0.18\textwidth,width=0.22\textwidth]{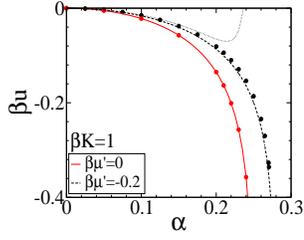}&
 \end{tabular}
 \end{center}
\caption{Internal energy (for indistinguishable particles) as a function of $\alpha$ for $\beta K=1$, 
and $\beta \mu'=0$ and $\beta \mu'=-0.2$.   The data points are from 
Monte Carlo simulation.  The solid line is for $\beta u$ in Eq. (\ref{eq:u}).  The dotted line incorporates 
the perturbative correction in Eq. (\ref{eq:du}).  The dashed line is for the variational approach.}
\label{fig:du} 
\end{figure}

Fig. (\ref{fig:du}) plots the data points for $\beta u$, for $\beta K=1$ and two values of the 
chemical potential, $\beta \mu'=0$ and $\beta \mu' =-0.2$, the former corresponding to 
infinite and the latter to a finite density.  The data points indicate that the reduced density
leads to higher internal energy.  The perturbative correction in Eq. (\ref{eq:du}) 
for the case $\beta \mu' =-0.2$ is shown as a dotted line. It accurately represents the simulated 
results for $\alpha<0.15$, then for $\alpha>0.15$ it becomes increasingly less accurate, and eventually 
diverges in the wrong direction as $\alpha\to 1/4$.  Because the perturbation approach 
breaks down, it cannot tell us anything about the value of $\beta u$ in a metastable region.  
%%%%%%%%%%%%%%%%%%%%%%
\graphicspath{{figures/}}
%%%%%%%%%%%%%%%%%%%%%%%

We next turn to a variational method.  We start by postulating a quadratic auxiliary Hamiltonian 
$$
H_{\Gamma}  =  \frac{K}{2} s^T \Gamma s, 
$$
where $\Gamma$ is a $N\times N$ matrix.  To keep things simple, it is assumed that $\Gamma$
has the same structure as the matrix $A$, and the only difference is that the coupling constant does
not correspond to a physical value $\alpha$ but is used as a variational parameter designated by 
$\alpha'$.  The partition function written in terms of the auxiliary ensemble is 
\be
\Xi = \Big\langle e^{-\frac{\beta K}{2} s^T (A-\Gamma) s - \beta \sum h(s_i)} \Big \rangle_{\! \Gamma}  \Xi_{\Gamma}.  
\ee
Then, using the Gibbs-Bogoliubov-Feynman inequality (GBF) \cite{Frydel15}, we get 
\be
\Xi \ge  e^{-\langle\frac{\beta K}{2} s^T (A-\Gamma) s - \beta \sum h(s_i)\rangle_{\Gamma} }  \Xi_{\Gamma}, 
\ee
and the quantity $\ln \Xi$ becomes  
\be
\ln \Xi \ge \ln \Xi_{\Gamma} -  \bigg\langle \frac{\beta K}{2} s^T (A-\Gamma) + N\beta h(s)\bigg\rangle_{\Gamma}.  
\ee
As the auxiliary system is Gaussian, the term in angular brackets can be evaluated, leading to 
\ba
\ln \Xi &\ge& \ln \Xi_{eff} = 
\ln \Xi_{\Gamma}+ \beta\mu' \frac{2N}{\pi}\sqrt{\frac{{\rm K}(16\alpha'^2)}{\beta K} } \nonumber\\
&+& N\bigg(\frac{1}{2}-\frac{1}{\pi}{\rm K}(16\alpha'^2) \bigg) \bigg(1-\frac{\alpha}{\alpha'} \bigg).
\label{eq:lnXi}
\ea

\begin{figure}[h] 
 \begin{center}
 \begin{tabular}{rrrr}
  \includegraphics[height=0.18\textwidth,width=0.22\textwidth]{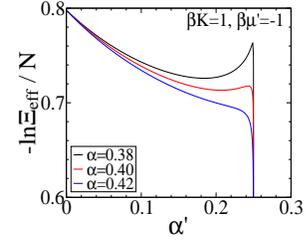}&
 \end{tabular}
 \end{center}
\caption{$-\ln\Xi$ as a function of a variational parameter $\alpha'$
for $\beta K=1$ and $\beta \mu'=-1$ (for indistinguishable particles), for three different 
values of $\alpha$.  }
\label{fig:Xi_eff} 
\end{figure}

Fig. (\ref{fig:Xi_eff}) plots $-\ln \Xi_{eff}/N$, where $\ln \Xi_{eff}$ is given in Eq. (\ref{eq:lnXi}), 
as a function of a variational parameter $\alpha'$.  Because the plots are for $\alpha>1/4$, 
the local minima in those plots correspond to metastable equilibriums.  The minimum disappears 
at around $\alpha\approx 0.42$, in which case the system spontaneously collapses.  
%as a function of $\alpha'$, where $\beta F_{eff}= -\ln \Xi_{eff}$ corresponds to a minimum.  
%%%%%%%%%%%%%%%%%%%%%%
\graphicspath{{figures/}}
%%%%%%%%%%%%%%%%%%%%%%%

The free energy of a metastable equilibrium corresponds to the function $-\ln\Xi_{eff}$ at a local  
minimum.  The internal energy is subsequently obtained from the definition in Eq. (\ref{eq:u0}).  
$\beta u$ obtained in this way is shown in Fig. (\ref{fig:du}) for the parameters $\beta K=1$
and $\beta\mu'=-0.2$ as a dashed line.  Comparison with the exact results indicates high 
degree of accuracy of the variational approach.  

%Encouraged by the accuracy of a variational approach, we use it next to predict the boundary 
%of the region of stability and its dependence on density.  
If we take the value of $\alpha$ where the local minimum of the function $-\ln\Xi_{eff}$
disappears, see Fig. (\ref{fig:Xi_eff}), 
to indicate the end of the stability region, we can use the variational method to 
obtain precise contours of the stability region.

\begin{figure}[h] 
 \begin{center}
 \begin{tabular}{rrrr}
  \includegraphics[height=0.18\textwidth,width=0.22\textwidth]{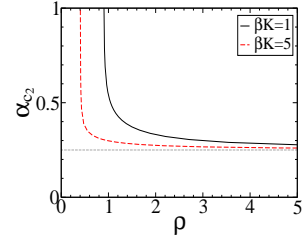}&
 \end{tabular}
 \end{center}
\caption{Boundaries of the metastable region as a function of the particle density for indistinguishable particles.  
Global minimum corresponds to $\alpha=1/4$ regardless of density.  The metastable region 
extends above this value and strongly depends on density.  }
\label{fig:alphac2} 
\end{figure}

Fig. (\ref{fig:alphac2}) plots such a boundary of the metastable region.  To make contact with 
the original particle system, we plot the results as a function of a particle density.  The density 
has been obtained from Eq. (\ref{eq:rho2b}) and within the variational framework is given by 
\be
\rho =  \sqrt{\frac{4{\rm K}(16{\alpha'}_0^2)} {\beta K \pi^2} } + \frac{2e^{2\beta \mu'}}{1-e^{2\beta \mu'}}, 
\ee
where ${\alpha'}_0$ corresponds to $\alpha'$ at a local minimum just as it is about to disappear.  
The results show drastic broadening of the metastable region as $\rho<1$.  This effect 
is even stronger for smaller $\beta K$.  
%%%%%%%%%%%%%%%%%%%%%%
\graphicspath{{figures/}}
%%%%%%%%%%%%%%%%%%%%%%%

For distinguishable particles we see the same type of general behavior and the emergence of the 
metastable region.  However, the application of the variational 
procedure is more complex as the function $h(s)$ is more difficult to handle.

\section{Conclusion}
\label{sec:con}

This work investigates thermodynamic collapse in a two-component lattice-gas system 
of penetrable particles on a square-lattice substrate.  Because particles are penetrable, 
there is no limit on how many particles occupy the same site, and the multiple occupation 
of a single site gives rise to different statistical mechanics, depending whether particles
are regarded as distinguishable or indistinguishable.  To facilitate analysis of the system, 
we transform the relevant partition function into the spin model with spins 
$s_i=0,\pm1,\pm2,\dots$.  In the limit $\rho\to\infty$, the system Hamiltonian recovers a 
simple quardatic form, so that the partition function corresponds to a discrete Gaussian 
model analyzed in the past in connection to interfaces and the roughening 
transition.  The difference between the Gaussian model used to study interfaces and 
the Gaussian model of penetrable particles lies in the sign of interactions between spins.  
Because the Gaussian model at the point of a collapse becomes isomorphic with 
the lattice Coulomb system, we check for the existence of a KT phase transition along
the line of the thermodynamic instability.  The presence of the KT transition itself 
does not affect the collapse transition, it might, however, affect the mechanism.  
%However, both simulation and our analytical 
%analysis do not reveal any indication of a phase transition.  We believe that the
%absence of a transition is attributed to the fact that sites in our model are multiply  
%occupied, so that the properties of the Coulomb lattice model with the occupation 
%number limited to one are not guaranteed to be reproduced in our system.  

To analyze the system for finite $\rho$ we employ a variational approximation
since for this situation the Hamiltonian is no longer harmonic.  
Both simulations and the approximation indicate the presence of a metastable 
equilibrium corresponding to a local minimum in the free energy.  The extent of the 
metastable region, furthermore strongly depends on density.  The metastable 
region vanishes at an infinite density, and diverges as density goes to zero.

\appendix 

\section{Selected relations of the Gaussian integral} 
\label{sec:A0}
The partition function of a continuous Gaussian model is a 
Gaussian integral, 
\be
\Xi = \int_{-\infty}^{\infty} ds_1\dots \int_{-\infty}^{\infty} d s_N\, e^{-\frac{1}{2}s^T B s} = 
\sqrt{\frac{(2\pi)^N}{\det { B}}}, 
\ee
where $B$ is the $N\times N$ square and symmetric matrix and $s=(s_1,\dots,s_N)$ is the $N$-dimensional 
vector.  

The probability that a spin $i$ has a value $s_i'$ can be obtained from the following 
definition 
\be
p(s'_i) = \langle \delta(s_i-s'_i) \rangle.  
\ee
Using the Fourier representation of a delta function, the relation above becomes 
\be
p(s'_i) = \frac{1}{2\pi}\int_{-\infty}^{\infty} dq\, \big\langle  e^{iq (s_i-s'_i)} \big\rangle, 
\ee
or, if we want to be more explicit 
\be
p(s'_i) = \int_{-\infty}^{\infty} \!\!\! dq\,  \frac{e^{-iq s'_i}}{2\pi \,\Xi}
\bigg[\int_{-\infty}^{\infty} \!\!\! ds_1\dots \int_{-\infty}^{\infty} \!\!\! d s_N\,  e^{iq s_i} e^{-\frac{1}{2}s^T B s}\bigg], 
\ee
where the integral inside the square brackets is the Gaussian integral with the linear term 
which after evaluation leads to 
\be
p(s'_i) =  \int_{-\infty}^{\infty} dq\,  \frac{e^{-iq s'_i}}{2\pi} e^{-\frac{1}{2}q^2 B_{ii}^{-1}}, 
\ee
which evaluates to 
\be
p(s_i) = e^{-\frac{s_i^2}{2B^{-1}_{ii}}}   \sqrt{\frac{1}{2\pi B^{-1}_{ii}}}, 
\label{eq:A3_2}
\ee
where $B^{-1}_{ii}$ is the element of the inverse matrix $B^{-1}$.  
%The fact that the distribution $p(s_i)$ is Gaussian is not surprising given the Gaussian form of the partition function.  
Using the distribution $p(s_i)$, the second moment of a spin $s_i$ is 
\be
\langle s_i^2 \rangle_g = B^{-1}_{ii}.  
\ee
We can also evaluate the average value of $|s_i|$, 
\be
\langle |s_i| \rangle_g =  \sqrt{\frac{2B^{-1}_{ii}}{\pi }}.  
\label{eq:A3_3}
\ee
%which shows
%\be
%\sqrt{\langle s_i^2 \rangle} > \langle |s_i| \rangle_g.  
%\ee

A similar procedure can be used to calculate a two spin distribution function 
\be
p(s'_i,s'_j) = \Big\langle \delta(s_i-s'_i) \delta(s_j-s'_j) \Big\rangle,
\ee
for $i\ne j$.  Using the Fourier representation of a delta function we get
\be
p(s'_i,s'_j) = \bigg(\frac{1}{2\pi}\bigg)^2 \!\! \int_{-\infty}^{\infty} \!\!\! dq_1\int_{-\infty}^{\infty} \!\!\! dq_2\, 
\Big\langle  e^{iq_1 (s_i-s'_i)} e^{iq_2 (s_j-s'_j)} \Big\rangle.  
\ee
If we follow similar steps taken to obtain $p(s'_i)$, we may obtain the expression for $p(s'_i,s'_j)$ 
which then allows us to calculate the spin-spin correlation function that evaluates to 
%\ba
%p(s'_i,s'_j) &=& \int_{-\infty}^{\infty} \!\!\! dq_1\int_{-\infty}^{\infty} \!\!\! dq_2\,  \frac{e^{-iq_1 s'_i}e^{-iq_2 s'_j}}{4\pi^2 \,\Xi}\nonumber\\
%&\times&
%\bigg[\int_{-\infty}^{\infty} \!\!\! ds_1\dots \int_{-\infty}^{\infty} \!\!\! d s_N\,  e^{iq_1 s_i}e^{iq_2 s_j} e^{-\frac{\beta}{2}s^T B s}\bigg], \nonumber\\
%\ea
%\ba
%p(s'_i,s'_j) &=& \int_{-\infty}^{\infty} \!\!\! dq_1\int_{-\infty}^{\infty} \!\!\! dq_2\,  \frac{e^{-iq_1 s'_i}e^{-iq_2 s'_j}}{4\pi^2 \,}\nonumber\\
%&\times&
%e^{-\frac{1}{2} q_1^2 B_{ii}^{-1} } e^{-\frac{1}{2} q_2^2 B_{ii}^{-1} }  e^{- q_1q_2 B_{ij}^{-1} } \nonumber\\
%\ea
%\be
%p(s'_i,s'_j) = e^{-\frac{1}{2} (s'_is'_i B_{ii}^{-1} + s'_js'_j B_{jj}^{-1} - 2s'_is'_j B_{ij}^{-1}) / (B^{-1}_{ii} B^{-1}_{jj} - B^{-1}_{ij}B^{-1}_{ij}) } \frac{1}{2\pi\sqrt{ B^{-1}_{ii} B^{-1}_{jj} - B^{-1}_{ij}B^{-1}_{ij} }}
%\ee
\be
\langle s_i s_j \rangle_g = B^{-1}_{ij}.  
\ee
%\be
%\langle |s_i s_j| \rangle = B^{-1}_{ij}.  
%\ee
Within the continuous Gaussian model, therefore, the inverse of the interaction matrix corresponds to the 
spin-spin correlation function.

\section{Matrix ${A}$ for the continuous Gaussian model} 
\label{sec:A1}

In this section we obtain the matrix ${A}$ 
of the continuous Gaussian model for an arbitrary dimension $d$.  
For the sake of concreteness, we assume the system size to be $L=4$, and in 
$d=1$ the system configuration can be represented with a vector 
%where the number of lattice sites is the same as the system size, $N=L$.  
%assume the system to be made up of four spins, $L=4$, 
$$
\begin{array}{|c|c|c|c|}
\hline s_1 & s_2 & s_3 & s_4 \\ \hline
\end{array}
$$
and the interaction matrix ${A}$ for the periodic boundary conditions is 
\be
{A}_{L} = 
\begin{bmatrix}
    1 & \alpha & 0 & \alpha  \\
    \alpha & 1 & \alpha  & 0 \\
    0 & \alpha & 1 & \alpha  \\
     \alpha & 0 & \alpha & 1 
\end{bmatrix}
     \label{eq:A1}
\ee
where the subscript $L$ denotes the matrix size $L\times L$.  The  
matrix is circulant, symmetric, and real valued.  Because only three elements are non-zero, 
the matrix, furthermore, is circulant tridiagonal.

In $d=2$, the spins of the system with size $L=4$ can be represented on a square grid as 
$$
\begin{array}{|c|c|c|c|}
\hline s_1 & s_2 & s_3 & s_4 \\ \hline
s_5 & s_6 & s_7 & s_8 \\ \hline
s_9 & s_{10} & s_{11} & s_{12 }\\ \hline
s_{13} & s_{14} & s_{15} & s_{16} \\ \hline
\end{array}
$$
and the resulting $A$ matrix for the nearest neighbor interactions is given by 
\begin{align}
{A}_{L^2} = 
\begin{bmatrix}
    \begin{array}{cccc|cccc|cccc|cccc}
    1 & \alpha & 0 & \alpha  & \alpha & 0 & 0 & 0 & 0 & 0 & 0 & 0 & \alpha & 0 & 0 & 0  \\
    \alpha & 1 & \alpha  & 0 & 0 & \alpha & 0 & 0 & 0 & 0 & 0 & 0& 0 & \alpha & 0 & 0  \\
    0 & \alpha & 1 & \alpha  & 0 & 0 & \alpha & 0 & 0 & 0 & 0 & 0& 0 & 0 & \alpha & 0  \\
     \alpha & 0 & \alpha & 1 & 0 & 0 & 0 & \alpha & 0 & 0 & 0 & 0& 0 & 0 & 0 & \alpha  \\ \hline
     \alpha & 0 & 0 & 0 & 1 & \alpha & 0 & \alpha & \alpha & 0 & 0 & 0 & 0 & 0 & 0 & 0 \\
    0 & \alpha & 0 & 0 & \alpha & 1 & \alpha  & 0 & 0 & \alpha & 0 & 0 & 0 & 0 & 0 & 0  \\
    0 & 0 & \alpha & 0 & 0 & \alpha & 1 & \alpha &  0 & 0 & \alpha & 0  & 0 & 0 & 0 & 0\\
    0 & 0 & 0 & \alpha & \alpha & 0 & \alpha & 1 &  0 & 0 & 0 & \alpha  & 0 & 0 & 0 & 0\\ \hline
    0 & 0 & 0 & 0 & \alpha & 0 & 0 & 0 &  1 & \alpha & 0 & \alpha  & \alpha & 0 & 0 & 0\\
    0 & 0 & 0 & 0 & 0 & \alpha & 0 & 0 &  \alpha & 1 & \alpha & 0 & 0 & \alpha & 0 & 0\\
    0 & 0 & 0 & 0 & 0 & 0 & \alpha & 0 &  0 & \alpha & 1 & \alpha  & 0 & 0 & \alpha & 0 \\
    0 & 0 & 0 & 0 & 0 & 0 & 0 & \alpha &  \alpha & 0 & \alpha & 1  & 0 & 0 & 0 & \alpha \\ \hline
    \alpha & 0 & 0 & 0 & 0 & 0 & 0 & 0 & \alpha & 0 & 0 & 0 &  1 & \alpha & 0 & \alpha\\
    0 & \alpha & 0 & 0 & 0 & 0 & 0 & 0 & 0 & \alpha & 0 & 0 &  \alpha & 1 & \alpha & 0\\
    0 & 0 & \alpha & 0 & 0 & 0 & 0 & 0 & 0 & 0 & \alpha & 0 &  0 & \alpha & 1 & \alpha\\
    0 & 0 & 0 & \alpha & 0 & 0 & 0 & 0 & 0 & 0 & 0 & \alpha &  \alpha & 0 & \alpha & 1\\
\end{array}
\end{bmatrix}
\label{eq:A1_2}
\end{align}
where the size of the matrix is $L^2\times L^2$.  If we subdivide the matrix into equally 
sized square blocks, we find three different sub-matrices.  The diagonal blocks are identical 
to the matrix in Eq. (\ref{eq:A1}).  The blocks adjacent to it are diagonal matrices with the 
diagonal element $\alpha$, and the remaining blocks are zero matrices.  The matrix 
$A_{L^2}$ can more conveniently be represented as a $L\times L$ matrix whose elements 
in turn are $L\times L$ matrices, 
\[
{A}_{L^2} = 
\begin{bmatrix}
    {A}_L & \alpha {I}_L & {0}_L & \alpha {I}_L  \\
    \alpha {I}_L & {A}_L & \alpha  {I}_L & {0}_L \\
    {0}_L & \alpha {I}_L & {A}_L & \alpha {I}_L \\
     \alpha {I}_L & {0}_L & \alpha  {I}_L & {A}_L 
\end{bmatrix}
\]
where ${I}_L$ is an identity and ${0}_L$ is a zero square matrix.  
The block representation of the matrix $A_{L^2}$ is a circulant matrix.

The block representation of the matrix ${A}_{L^3}$ in $d=3$ is 
\[
{A}_{L^3} = 
\begin{bmatrix}
    {A}_{L^2} & \alpha {I}_{L^2} & {0}_{L^2} & \alpha {I}_{L^2}  \\
    \alpha {I}_{L^2} & {A}_{L^2} & \alpha  {I}_{L^2} & {0}_{L^2} \\
    {0}_{L^2} & \alpha {I}_{L^2} & {A}_{L^2} & \alpha {I}_{L^2} \\
     \alpha {I}_{L^2} & {0}_{L^2} & \alpha {I}_{L^2} & {A}_{L^2}
\end{bmatrix}.
\]
%where the actual size of the matrix is $L^3\times L^3$ and the
%size of the constituent block matrices is $L^2\times L^2$.

%For a lattice system in an arbitrary dimension $d$, the matrix ${\bf A}$, employing 
%the block notation, should be
%\[
%{\bf A}_{L^d} = 
%\begin{bmatrix}
%    {\bf A}_{L^{d-1}} & \alpha {\bf I}_{L^{d-1}} & {\bf 0}_{L^{d-1}} & \alpha {\bf I}_{L^{d-1}}  \\
%    \alpha {\bf I}_{L^{d-1}} & {\bf A}_{L^{d-1}} & \alpha  {\bf I}_{L^{d-1}} & {\bf 0}_{L^{d-1}} \\
%    {\bf 0}_{L^{d-1}} & \alpha {\bf I}_{L^{d-1}} & {\bf A}_{L^{d-1}} & \alpha {\bf I}_{L^{d-1}} \\
%     \alpha {\bf I}_{L^{d-1}} & {\bf 0}_{L^{d-1}} & \alpha {\bf I}_{L^{d-1}} & {\bf A}_{L^{d-1}}
%\end{bmatrix}
%\]

\section{Onsager's exact solution of the Ising model}
\label{sec:A2}
For an antiferromagnetic Ising model with the Hamiltonian 
\be
H =  J \sum_{nn} s_i s_j, 
\ee
the free energy for a square-lattice geometry is given by \cite{Onsager44} 
\ba
\frac{\beta F}{N} &=& - \ln[2\cosh(2\beta J)] \nonumber\\
&-& \frac{1}{8\pi^2} \!\! \int_0^{2\pi} \!\!\! dq_1 \! \int_0^{2\pi} \!\!\! dq_2\, \ln\big[1 + 2\alpha \cos q_1 +  2\alpha \cos q_2\big], \nonumber\\
\ea
where $N=L^2$ and 
\be
\alpha = \frac{1}{2} \frac{\sinh(2\beta J)}{\cosh^2(2\beta J)}
\label{eq:A2_3}
\ee
is the function of the interaction strength.  
In view of the similarity of the integral term to that in Eq. (\ref{eq:I}), we may write 
\be
\frac{\beta F}{N} = -\ln[2\cosh(2\beta J)] + \frac{1}{4}\sum_{k=1}^{\infty} \frac{\alpha^{2k}}{k} \frac{(2k)!^2}{k!^4},
\ee
(recently, a similar expression, in terms of $_4F_3$ hypergeometric function, 
has been obtained in \cite{Viswanathan15}), and the corresponding partition function can be written as 
\be
\Xi_{IS} = [2\cosh(2\beta J)]^N \exp\bigg[-\frac{N}{4}\sum_{k=1}^{\infty} \frac{\alpha^{2k}}{k} \frac{(2k)!^2}{k!^4}\bigg].  
\ee
Knowing that the series in the above expression has a convergence interval $|\alpha|\le 1/4$, 
the phase transition must occur at $\alpha_c=1/4$, on the edge of the stability 
region.  Using Eq. (\ref{eq:A2_3}), this corresponds to 
\be
\frac{1}{2} \frac{\sinh(2\beta J_c)}{\cosh^2(2\beta J_c)} = \frac{1}{4}, 
\ee
which yields $\beta J_c = \ln(1+\sqrt{2})/2$.  (A number of interesting results for the Ising model 
in two-dimension based on series approach can be found in \cite{Perk11}).

\section{Connection with the Chui-Weeks model}
\label{sec:A3}

The DG surface model of Chui and Weeks \cite{Chui76} can be represented by the following Hamiltonian, 
\be
H_{} =  \frac{J}{2} \sum_{j\ne i}^N  \epsilon_{ij} (s_i-s_j)^2 + 4h J \sum_{i=1}^{N} s_i^2, 
\ee
where $\epsilon_{ij}=1$ if two spins are the nearest neighbors and $\epsilon_{ij}=0$ otherwise.  
To connect the Chui-Weeks system to the quadratic Hamiltonian in Eq. (\ref{eq:H_infty_B})
corresponding to the limit $\rho\to\infty$, 
\be
H_{\infty} =  \frac{K}{2}  \sum_{j\ne i}^N  \alpha \epsilon_{ij} s_i s_j + \frac{K}{2} \sum_{i=1}^{N} s_i^2, 
\ee
we rewrite the above expression using $2 s_i s_j =  s_i^2 + s_j^2 - (s_i - s_j)^2$.  This leads to 
\be
H_{\infty} =  -\frac{K}{4} \sum_{j\ne i}^N  \alpha \epsilon_{ij} (s_i-s_j)^2
+ \frac{K}{2}\alpha \sum_{j\ne i}^N  \epsilon_{ij} s_i^2
+ \frac{K}{2}  \sum_{i=1}^{N} s_i^2.  
\ee
Because in 2D there are four neighbors, this simplifies to 
\be
H_{\infty} =  -\frac{\alpha K}{4}  \sum_{j\ne i}^N  \epsilon_{ij} (s_i-s_j)^2 + \frac{K}{2} (1+4\alpha)\sum_{i=1}^{N} s_i^2.  
\ee
By comparing the parameters of the Chui-Weeks model with the model governed by the Hamiltonian in 
Eq. (\ref{eq:H_infty_B}) we get 
$$
J = -\frac{\alpha K}{2},~~~~
h  = -\frac{1+4\alpha}{4\alpha}.  
$$
For $\alpha=-1/4$, our model corresponds to the case $h=0$, for which it becomes isomorphic
with the lattice Coulomb model.

\begin{acknowledgments}
D.F. acknowledges financial support from FONDECYT through grant number 1201192.  
D.F. thanks the University of Tel Aviv for invitation under the program the ``Visiting 
Scholar of The School of Chemistry'', and the hospitality of Haim Diamant and David 
Andelman, during which a part of this manuscript 
was completed.  All computations were done on the UFTSM computer cluster 
managed by Yuri Ivanov.     
%D.F. acknowledges financial support of the Agencia Nacional de Investigación y Desarrollo (ANID), 
%018/FONDECYT/39, Grant Number 1201192.  
%via the ``Programa 3: Apoyo a la Instalaci\'on en Investigaci\'on''.  
\end{acknowledgments}

%------------------------------------------------
% References
%------------------------------------------------

%------------------------------------------------

\end{document}